\documentclass[journal,twoside,web]{ieeecolor}
\usepackage{tmi}

\usepackage{hyperref} 
\usepackage{times}
\usepackage{epsfig}
\usepackage{graphicx}
\usepackage{pifont}
\usepackage{epsfig}
\usepackage{booktabs,multirow}
\usepackage{graphics}
\usepackage{threeparttable}
\usepackage{cite}
\usepackage{amsmath,amssymb,amsfonts}
\usepackage{algorithmic}
\usepackage{textcomp}
\usepackage{threeparttable}
\usepackage{color}
\usepackage[normalem]{ulem}
\usepackage{multirow}
\usepackage{float}
\usepackage{amsfonts}
\usepackage{bbm}
\usepackage{multirow}
\usepackage{subfig}
\usepackage{array}
\usepackage[table]{xcolor}
\usepackage{colortbl}
\definecolor{bblue}{rgb}{0,150,230}
\definecolor{mygray}{gray}{.9}
\definecolor{myy}{RGB}{126,95,0}
\usepackage{bm}

\newcommand{\eg}[1]{\textit{e.g.,}}
\newcommand{\ie}[1]{\textit{i.e.,}}

\newcommand{\figref}[1]{Fig.\!~\ref{#1}}

\makeatletter
\newcommand{\thickhline}{%
	\noalign {\ifnum 0=`}\fi \hrule height 1pt
	\futurelet \reserved@a \@xhline
}
\makeatother

\usepackage{cleveref}
\crefname{section}{}{§§}
\Crefname{section}{}{§§}

\usepackage{caption}
\def\BibTeX{{\rm B\kern-.05em{\sc i\kern-.025em b}\kern-.08em
    T\kern-.1667em\lower.7ex\hbox{E}\kern-.125emX}}
    
\begin{document}

\title{Multi-Modal Transformer for Accelerated \\MR Imaging}

\author{Chun-Mei Feng, Yunlu Yan, Geng Chen, Yong Xu, \IEEEmembership{Senior Member, IEEE}, \\ Ying Hu, Ling Shao, \IEEEmembership{Fellow, IEEE}, and Huazhu Fu, \IEEEmembership{Senior Member, IEEE},  
\thanks{This work was supported by the National Key R\&D Program of China [2018AAA0100100] and AME Programmatic Fund (A20H4b0141).}
\thanks{C.-M.~Feng, Y.~Yan, Y.~Hu and Y.~Xu are with the Shenzhen Key Laboratory of Visual Object Detection and Recognition, Harbin Institute of Technology (Shenzhen), 518055, China.~(Email: strawberry.feng0304@gmail.com; yongxu@ymail.com).}
\thanks{G. Chen is with National Engineering Laboratory for Integrated Aero-Space-Ground-Ocean Big Data Application Technology, School of Computer Science and Engineering, Northwestern Polytechnical University, Xi'an 710072, China (e-mail: geng.chen.cs@gmail.com).}
\thanks{L.~Shao is with the Terminus Group, China. (Email: ling.shao@ieee.org).}
\thanks{H.~Fu is with the Institute of High Performance Computing, A*STAR, Singapore 138632. (E-mail: hzfu@ieee.org).}
\thanks{Corresponding author: \textit{Yong Xu and Huazhu Fu}.}
\thanks{C.-M.~Feng and Y.~Yan are contributed equally to this work.}
}
\maketitle

\begin{abstract}

Accelerated multi-modal magnetic resonance (MR) imaging is a new and effective solution for fast MR imaging, providing superior performance in restoring the target modality from its undersampled counterpart with guidance from an auxiliary modality. 
However, existing works simply combine the auxiliary modality as prior information, lacking in-depth investigations on the potential mechanisms for fusing different modalities.
Further, they usually rely on the convolutional neural networks (CNNs), which is limited by the intrinsic locality in capturing the long-distance dependency.
To this end, we propose a multi-modal transformer (MTrans), which is capable of transferring multi-scale features from the target modality to the auxiliary modality, for accelerated MR imaging.
To capture deep multi-modal information, our MTrans utilizes an improved multi-head attention mechanism, named cross attention module, which absorbs features from the auxiliary modality that contribute to the target modality. 
Our framework provides three appealing benefits: (i) Our MTrans use an improved transformers for multi-modal MR imaging, affording more global information compared with existing CNN-based methods. (ii) A new cross attention module is proposed to exploit the useful information in each modality at different scales. The small patch in the target modality aims to keep more fine details, the large patch in the auxiliary modality aims to obtain high-level context features from the larger region and supplement the target modality effectively. (iii) We evaluate MTrans with various accelerated multi-modal MR imaging tasks, \eg, MR image reconstruction and super-resolution, where MTrans outperforms state-of-the-art methods on fastMRI and real-world clinical datasets.

\end{abstract}

\begin{IEEEkeywords}
MR imaging, multi-modal, reconstruction, super-resolution.
\end{IEEEkeywords}

\section{Introduction}

Magnetic resonance (MR) imaging is rapidly becoming the dominant technique for image-guided adaptive radiotherapy because it offers better soft tissue contrast than computed tomography (CT), while avoiding radiation exposure. However, due to the physical nature of the MR imaging procedure, the scanning time can take up to tens of minutes long, which seriously affects the patient experience and leads to high costs. Therefore, accelerated MR imaging has become a hot research topic, where reconstructing images from undersampled $k$-space measurements is a standard strategy. However, the aliasing artifacts caused by insufficient sampling often affect the clinical diagnosis. Therefore, the recovery of high-quality images from undersampled $k$-space measurements is the ultimate goal when accelerating MR imaging. Currently, mainstream methods for this include MR image reconstruction and super-resolution (SR). The former aims to remove the aliasing artifacts caused by undersampling~\cite{zhang2015exponential,chen2019model,aggarwal2018modl,Wang2020NMI,feng2021specificity}, while the latter enhances the image resolution~\cite{chen2018efficient,chaudhari2018super,mcdonagh2017context,feng2021exploring,feng2021multiq}.

\begin{figure}[!t]
\centering
  \includegraphics[width=0.48\textwidth]{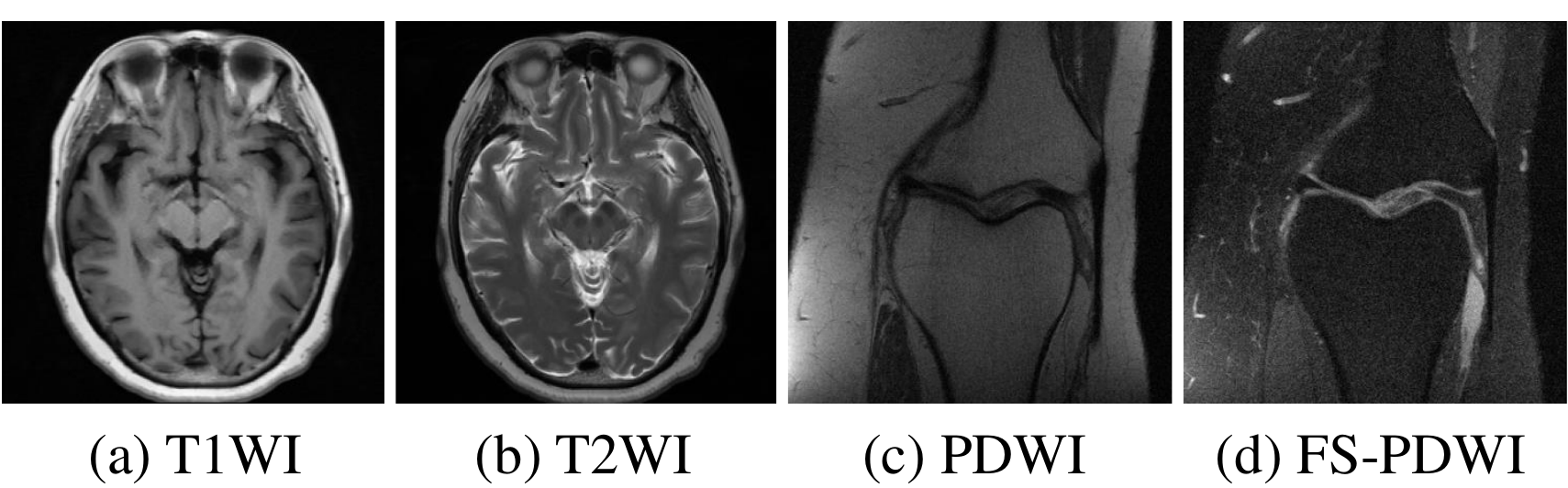}
  \caption{Examples of four different modalities. Images (a) and (b) are a T1WI and T2WI from the same subject of a real-world clinical dataset. Images (c) and (d) are a PDWI and FS-PDWI from the same subject of fastMRI. Different modalities from the same subject have inter-modality consistent structure.} 
  \label{fig1} 
\end{figure}

In practice, according to different acquisition sequences, the scanner usually acquires MR images with different modalities at the same time to meet the medical diagnosis. For the same subject, these modalities often have inter-modality consistent information and modality-specific information~\cite{xiang2018deep}.
In addition, the acquisition procedures of the different modalities vary.
For example, T1 and T2 weighted images (T1WIs and T2WIs), as well as proton density and fat-suppressed proton density weighted images (PDWIs and FS-PDWIs), are two pairs of images with complementary structures.
As shown in~\figref{fig1}, (a) and (b) are a pair of T1W and T2W brain MR images from the same subject of a real-world clinical dataset, where (a) provides morphological and structural information, and (b) shows edema and inflammation. Following~\cite{xuan2020learning}, we filter out pairs of PDW and FS-PDW images from fastMRI (currently the largest available database for raw MR images), as shown in~\figref{fig1} (c) and (d).
PDWIs provide structural information for articular cartilage, while FS-PDWIs can inhibit fat signals and highlight the structural contrast between cartilage ligaments and other tissues~\cite{chen2015accuracy}.
Generally, due to the physical characteristics of the MR imaging, T1WIs are more easily acquired than T2WIs since they require a shorter repetition time (TR) and echo time (TE)~\cite{xiang2018deep}. Especially in the same imaging sequence, the acquisition time of T2WI is usually longer than that of T1WI due to the longer TR time of T2, \eg, the acquisition time of T2SE is longer than T1SE~\cite{xiang2018deep}. 
Similarly, PDWIs require a shorter scan time than FS-PDWIs.
In clinical settings, since different modalities can provide different information, we usually acquire multiple modalities simultaneously to facilitate the comprehensive diagnosis of disease.
Therefore, we can use relatively easy-to-obtain modalities as supplementary information to guide and accelerate target modalities that are acquired with slower imaging speeds.
To this end, the overall acquisition time and difficulty will be greatly reduced, which is the practical significance of our work.

From the auxiliary modality (with faster imaging speed) to help the target modality (with slower imaging speed) to obtain a high-quality image is actually a process of accelerated imaging, which has been verified by the previous works~\cite{xiang2018deep,dar2020prior,lyu2020multi,zeng2018simultaneous,feng2021multii}. For example, compressed sensing (CS), Bayesian learning, dictionary learning, and graph representation theory, have been employed to accelerate multi-modal MR imaging~\cite{bilgic2011multi,song2019coupled,lai2017sparse}. Bilgic \textit{et al.}~introduced the Joint VC (JVC) technique to the GRAPPA framework for multi-modal MRI reconstruction~\cite{bilgic2018improving}.
Gong \textit{et al.}~used the shareable information among multi-modal images to accelerate MR imaging~\cite{gong2015promise}. More recently, deep learning has become the focus of multi-modal MR imaging studies~\cite{sun2019deep,dar2020prior}. For example, Dar \textit{et al.}~added an auxiliary modality as prior information into the generator of a generative adversarial network (GAN)~\cite{dar2020prior}. Liu \textit{et al.}~used an iterative network to make the shareable information among multi-modal images for accelerated MR imaging~\cite{liu2021regularization}.
Zhang \textit{et al.}~applied the temporal feature fusion block based on ADMM to achieve multi-modal MRI reconstruction~\cite{zhang2021temporal}. Lyu \textit{et al.}~concatenated the two modalities at the feature level of smaller size~\cite{lyu2020multi}. In addition, different MR image modalities have modality-specific appearances under different intensity distributions. \textit{Thus, how to effectively fuse the two modalities is an inherent problem in multi-modal MR imaging, which needs to be resolved.}

On the other hand, convolutional neural networks (CNNs) struggle to fully capture global knowledge due to its intrinsic locality of convolution operations, while transformers can learn global information by modeling  long-range dependency. Benefiting from this, transformers have recently achieved state-of-the-art performance on a variety of computer vision tasks~\cite{Khan2021}. For example, the Vision Transformer (ViT) divides images into small patches and uses a transformer to model the correlation between them as sequences, achieving satisfactory results in image classification~\cite{dosovitskiy2020image}. The Detection Transformer (DETR) formulates target detection as an ensemble prediction task with the help of a transformer~\cite{carion2020end}. Transformers have also been used in medical imaging tasks. For example, transformers incorporated into UNet have been employed for medical image segmentation~\cite{chen2021transunet,wang2021transbts}. Korkmaz \textit{et al.}~used the cross attention module to capture interactions between latent variables and image features with a deep adversarial network to solve the unsupervised MRI reconstruction problem~\cite{korkmaz2021unsupervised}. Feng \textit{et al.}~applied the task transform module, which is evolved from self-attention module of CNN, to transmit and share the representations between multiple tasks~\cite{feng2021task}.

Although transformers have been applied to MR image reconstruction, these efforts are focused on a single modality in the basic transformer framework, ignoring the correlation between different modalities, especially in different scales.
To this end, in this work, we investigate how to design a powerful transformer model capable of learning multi-modal representations to enhance various accelerated MR imaging tasks. We propose a multi-modal transformer (MTrans) to fuse the informative features from MR imaging scans of different modalities based on a multi-modal transformer.
Our method utilizes multi-scale patches generated by the two-branch transformer to represent different modalities, and merges them to complement each other.
Another key contribution of our work is to develop a feature fusion strategy for multi-modal MR imaging transformers, which, to the best of our knowledge, has not yet been investigated.
This is achieved with our effective multi-modal cross attention modules, each of which takes the features from the other branch as keys and values, employing then for effectively querying, and obtain useful information from the other modality.
In addition, the multi-scale patches for two branches can capture high-level context features and local details to complement each other.

Overall, the main contributions of our work are as follows:
\begin{itemize}
\item {We propose a novel transformer architecture, named MTrans, to accelerate multi-modal MR imaging. Benefited from the advantage of transformer, our work is able to capture rich global knowledge compared with the existing CNN-based methods\footnote{Code is available at: {\color{red}{\href{https://github.com/chunmeifeng/MTrans}{https://github.com/chunmeifeng/MTrans}.}}}.}
\item {We introduce a cross attention module to effectively extract useful information in each branch and then combine the features from multiple scales to afford both high-level context features from the larger region and local details to complement each other.}
\item {We evaluate our method on two fast MR imaging tasks, \eg, image reconstruction and SR, on fastMRI and a raw MR image datasets. The results show that our method is superior to other multi-modal MR imaging models in both qualitative and quantitative evaluation.}
\end{itemize}

\section{Related Work}\label{sec:related_work}

\subsection{Deep Learning for Accelerated MR Imaging}
Image reconstruction and SR techniques can improve image quality without changing the MR image acquisition hardware, therefore they have been widely used for accelerated MR imaging. Traditional technologies such as CS~\cite{haldar2010compressed}, low rank~\cite{shi2020spectral,pramanik2020deep,haldar2016p,he2016accelerated}, and dictionary learning~\cite{bhatia2014super,liu2020highly} have made progress in this task. More recently, deep learning has been widely used. Compared with traditional algorithms, which rely on the prior information of data, deep learning can make full use of the inherent characteristics of images contained in a large amount of training data~\cite{wang2016accelerating}. For example, Yang \textit{et al.}~proposed a model-based unrolling method by applying the alternating direction method of multipliers (ADMM) algorithm to optimize the architecture~\cite{4}. MoDL then introduced another model-based unrolling method combining prior regularization~\cite{aggarwal2018modl}. More recently, end-to-end approaches have shown advantages in accelerated MR imaging~\cite{wang_first}. Jin \textit{et al.}~applied UNet to capture spatial information for inverse problems related to MR imaging~\cite{jin2017deep}. A 3D CNN with a residual architecture was used to generate high-quality MR image scans of knees in~\cite{chaudhari2018super}. Chen \textit{et al.}~recovered high-quality image details from a densely connected SR network~\cite{chen2018brain}. Zhu \textit{et al.}~effectively estimated the mappings by manifold approximation (AUTOMAP) for MR image reconstruction~\cite{zhu2018image}. For example, Kim \textit{et al.}~used auto-calibrated recurrent and scan-specific deep networks to accelerate MR imaging from $k$-space data~\cite{kim2019loraki,akccakaya2019scan}. Han \textit{et al.}~used the domain adaptation techniques to remove the artifacts from the undersampled images~\cite{han2018deep}. Lee \textit{et al.}~\cite{lee2018deep} trained the magnitude and phase of MR image data separately and fused them to generate the output image. To address the shortcomings of CNNs in calculating complex MR image numbers, we recently proposed Dual-OctConv to deal with complex components at various spatial frequencies for accelerated parallel MR imaging~\cite{feng2021DualOctConv,feng2021DONet}. This method not only considers the computational relationship between the real and imaginary parts, but also captures the characteristics of different spatial frequencies. Finally, inspired by the prominent use of GANs in natural image synthesis, many works have used GANs with an adversarial and perceptual loss to generate high-quality MR images~\cite{quan2018compressed, yang2017dagan, 15,lyu2018super}. 
The data consistency layer plays an important role in MR imaging to keep the reconstructed image consistent with the original image in the $k$-space~\cite{schlemper2017deep, zheng2019cascaded}. Additionally, hybrid domain learning schemes have been used to recover data from both the $k$-space and image domain~\cite{eo2018kiki}. However, these methods are all based on a single-modal CNN architecture. In contrast, our method is a multi-modal fusion approach based on the transformer architecture.

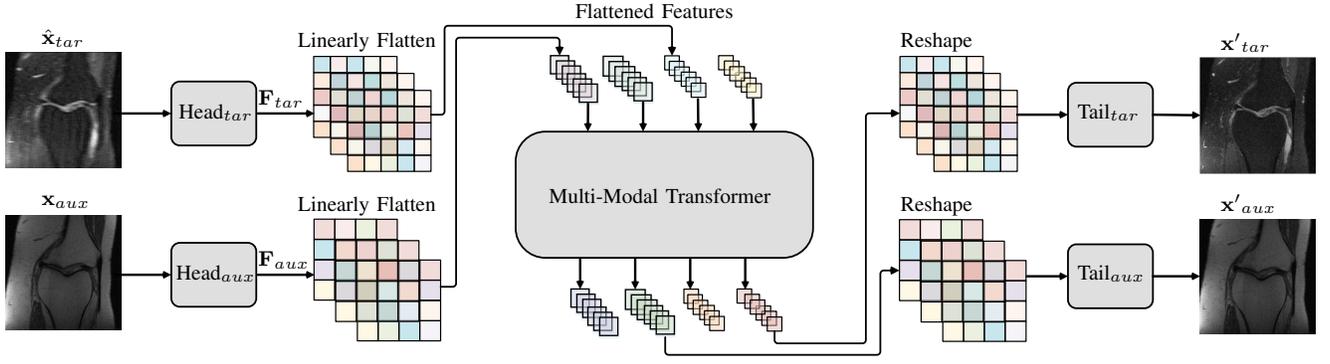
\begin{figure*}[t]
\centering
  \includegraphics[width=0.96\textwidth]{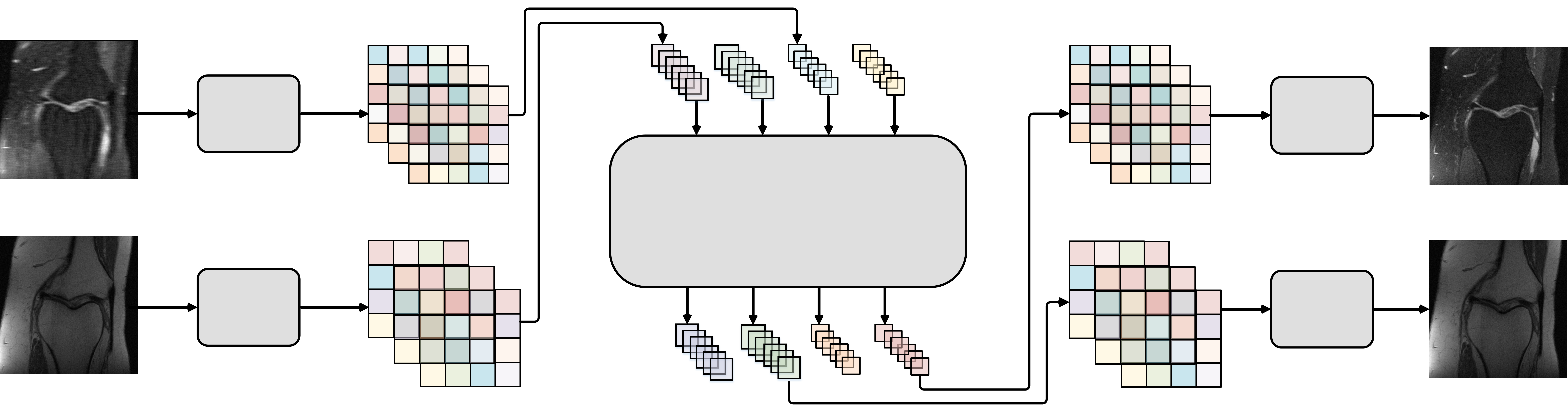}
  \put(-430,92){\footnotesize Head$_{tar}$}
  \put(-431,31){\footnotesize Head$_{aux}$}
  \put(-90,92){\footnotesize Tail$_{tar}$}
  \put(-90,31){\footnotesize Tail$_{aux}$}
  \put(-290,60){\footnotesize Multi-Modal Transformer}
  \put(-280,130){\footnotesize Flattened Features}
  \put(-385,119){\footnotesize Linearly Flatten}
  \put(-385,57){\footnotesize Linearly Flatten}
  \put(-157,119){\footnotesize Reshape}
  \put(-157,57){\footnotesize Reshape}
  \put(-482,120){\footnotesize ${\hat{\mathbf{x}}}_{tar}$}
  \put(-482,59){\footnotesize ${\mathbf{x}}_{aux}$}
  \put(-36,118){\footnotesize ${\mathbf{x}'}_{tar}$}
  \put(-36,57){\footnotesize ${\mathbf{x}'}_{aux}$} 
  \put(-400,97){\footnotesize $\mathbf{F}_{tar}$} 
  \put(-400,36){\footnotesize $\mathbf{F}_{aux}$} 
  \caption{\small \textbf{An illustration of the proposed MTrans framework}. Our architecture consists of two branches, \eg, a target branch and an auxiliary branch. This in turn are each divided into three components: two \textbf{heads} for extracting modality-specific features of different scales from the multi-modal input images (\eg, the fully sampled auxiliary modality image ${\mathbf{x}}_{aux}$ of large patch size with target zero filling ${\hat{\mathbf{x}}}_{tar}$ or the LR image ${\tilde{\mathbf{x}}}_{tar}$); a \textbf{multi-modal transformer} for aggregating the different modalities; and two \textbf{tails} in the target modality for mapping the features into restored images.}
  \label{fig2} 
\end{figure*}

\subsection{Multi-Modal Medical Image Representation}
Multi-modal fusion allows multiple modalities to be combined in a new space by taking advantage of complementarities between data, which is more robust than using any single modality as input. Recently, multi-modal technology has also been used widely in medical imaging~\cite{xiang2018deep,sun2019deep,dar2020prior,lyu2020multi,zeng2018simultaneous,zhou2019hyper}. For example, a hybrid-fusion network was designed for multi-modal MR image synthesis~\cite{zhou2020hi}. Xiang and Dar \textit{et al.}~simply concatenated the two modalities as input to guide the reconstruction and SR of the target modality~\cite{xiang2018deep,dar2020prior}. Sun \textit{et al.}~sent the different modalities into the network together to restore them simultaneously~\cite{sun2019deep}. For the MR image SR task, Lyu \textit{et al.}~concatenated the two modalities on the features of smaller size~\cite{lyu2020multi}, while Zheng \textit{et al.}~concatenated them on the original image size~\cite{zeng2018simultaneous}. Dai \textit{et al.}~used the transformer to establish long-range dependencies from the decomposed multi-modal image patches for MRI classification~\cite{dai2021transmed}. Li \textit{et al.}~integrated tensor modeling and deep learning for solving the multi-modal MR image processing problems~\cite{LimultiISMRM}. However, these existing multi-modal methods simply add the auxiliary modality as prior information for the target modality to improve the image quality; fusing the two modalities has not been explored~\cite{sun2019deep,xiang2018deep,dar2020prior,lyu2020multi}. 

\section{Overview of Accelerated MR Imaging}\label{sec:fast}
Let $\mathbf{y}$ represent the complex-valued, fully sampled $k$-space acquired from the MR image scanner. We can obtain the corresponding fully sampled image by $\mathbf{x}=\mathcal{F}^{-1}(\mathbf{y})$, where $\mathcal{F}^{-1}$ is an inverse 2D fast Fourier transform (FFT). In clinical practice, as only the magnitude images are visible, hospitals usually retain these for medical diagnosis. However, in this work, all data (\eg, the zero-filled image for reconstruction and LR image for SR) are obtained from real MR image $k$-space data to explore the effectiveness of accelerated MR imaging. This is an important point that has been neglected by current fast multi-modal MR imaging methods.
In this work, we consider two kinds of accelerate MR imaging techniques, including (i) reconstructing a clear image from an image corrupted by aliasing artifacts (undersampled image) and (ii) restoring an SR image from a degraded image.

\subsection{Accelerating MR Imaging by Reconstruction}\label{sec:rec}
Let $M$ be the binary mask operator. We can obtain the undersampled $k$-space data by $\hat{\mathbf{y}}= M \odot \mathbf{y}$, where $\odot$ denotes element-wise multiplication. In this work, we use random masks with 6$\times$ acceleration to select a subset of the $k$-space points. Accordingly, the zero-filled image can be obtained by $\hat{\mathbf{x}}=\mathcal{F}^{-1}(\hat{\mathbf{y}})$.
Different from current efforts, which address the task by directly restoring ${\mathbf{x}'}$ from $\hat{\mathbf{y}}$ or $\hat{\mathbf{x}}$, we introduce an image from an additional modality with the same structural information to restore the target modality.

\subsection{Accelerating MR Imaging by Super-Resolution}\label{sec:sr}
The training phases of previous MR image SR methods usually add Gaussian blurs to the downsampled amplitude image to obtain an LR image~\cite{pham2017brain}. However, simply reducing the image size in the image domain contradicts the actual MR image acquisition process. Following~\cite{chen2018efficient}, we first truncate the outer part of the fully sampled $k$-space $\mathbf{y}$ by a desired factor to degrade the resolution, and then apply $\mathcal{F}$ to obtain the degraded LR image $\tilde{{\mathbf{x}}}$. This better mimics the real image acquisition process and avoids checkerboard artifacts. 


\section{Proposed MTrans Architecture}\label{sec:proposed}
In our MTrans, the image patches are processed into a series of linearly embedded sequences to create a dual-branch structure. As shown in~\figref{fig2}, the overall architecture of our MTrans consists of three components. Specifically, two heads are employed for extracting modality-specific features of different scales from the multi-modal input images (\eg, a fully sampled auxiliary modality image of large size with target zero-filling or LR image of small size); a multi-modal transformer is established for aggregating the different modalities, where the module uses the feature of the current branch as the query to exchange information with the other branch; and two tails are used for mapping the features into restored images. Note that, the inputs from different modalities are divided into image patches of different sizes. This enables local details to be extracted, while also capturing high-level context features from larger regions to supplement the target modality. The main goal of the multi-modal transformer is to integrate multi-modal images at different scales. We will next introduce our architecture in detail.




\subsection{Heads}
To extract modality-specific features from different modalities, two branches with different heads (\eg, Head$_{aux}$ for the auxiliary modality and Head$_{tar}$ for the target modality) are used, each of which consists of three 3$\times$3 convolutional layers. The ground truth of the auxiliary modality ${\mathbf{x}}_{aux}\in \mathbb{R}^{1 \times H \times W}$ is sent to Head$_{aux}$ to generate an auxiliary feature map $\mathbf{F}_{aux} \in \mathbb{R}^{C \times H \times W}$, where $C$ is the number of channels, and $H$ and $W$ are the height and weight of the feature maps. For the MR image reconstruction task, we send the zero-filled image $\hat{\mathbf{x}}_{tar}$ of the target modality to Head$_{tar}$ to generate a feature map $\mathbf{F}_{tar} \in \mathbb{R}^{C \times H \times W}$, while the LR image ${\tilde{\mathbf{x}}}_{tar}$ is used to generate a feature map $\mathbf{F}_{tar} \in \mathbb{R}^{C \times \frac{H}{s}\!\times\!\frac{W}{s}}$ ($s$ is the scale factor) for the SR task. 

\subsection{Multi-modal Transformer}

\begin{figure*}[!t]
\centering
  \includegraphics[width=1\textwidth]{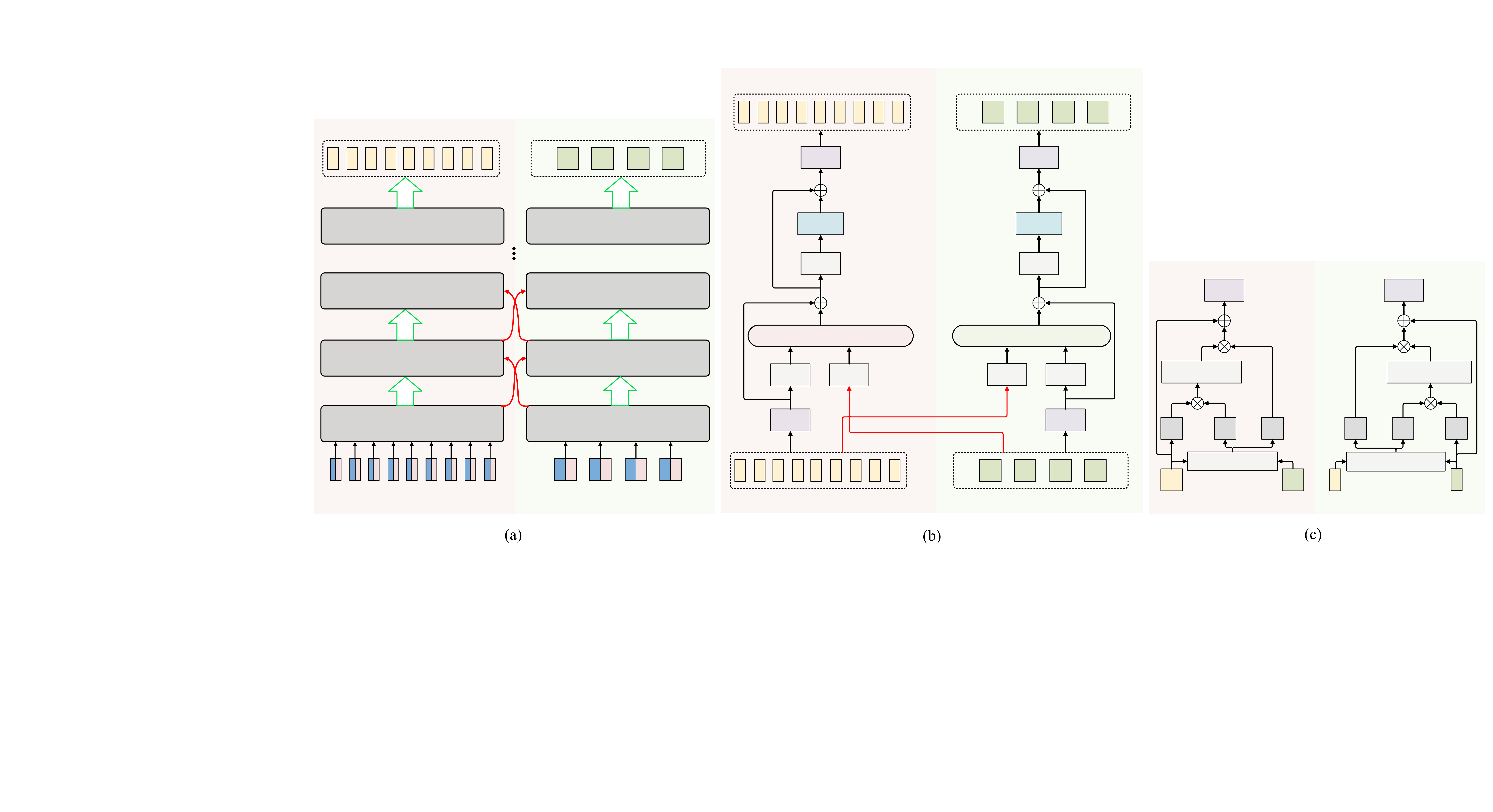}
  \small\put(-500,138){\footnotesize CT Encoder$_{tar_{N}}$}
  \put(-500,110){\footnotesize CT Encoder$_{tar3}$}
  \put(-500,81){\footnotesize CT Encoder$_{tar2}$}
  \put(-500,52){\footnotesize CT Encoder$_{tar1}$}
  \put(-410,138){\footnotesize CT Encoder$_{aux_{N}}$}
  \put(-410,110){\footnotesize CT Encoder$_{aux3}$}
  \put(-410,81){\footnotesize CT Encoder$_{aux2}$}
  \put(-410,52){\footnotesize CT Encoder$_{aux1}$}
  \put(-500,18){\footnotesize Target Branch}
  \put(-410,18){\footnotesize Auxiliary Branch}
  \put(-432,30){\footnotesize $\mathbf z^{0}_{tar}$}
  \put(-352,30){\footnotesize $\mathbf z^{0}_{aux}$}
  \put(-480,182){\footnotesize $\mathbf z^{l}_{tar}$} 
  \put(-390,182){\footnotesize $\mathbf z^{l}_{aux}$} %
  \put(-311,52){\footnotesize $\mathbf{LP}$}
  \put(-191,52){\footnotesize $\mathbf{LP}$}
  \put(-311,72){\footnotesize $\mathbf{LN}$}
  \put(-286,72){\footnotesize $\mathbf{LN}$}
  \put(-216,72){\footnotesize $\mathbf{LN}$}  
  \put(-191,72){\footnotesize $\mathbf{LN}$}   
  \put(-232,90){\footnotesize Cross Attention$_{aux}$}  
  \put(-320,90){\footnotesize Cross Attention$_{tar}$}  
  \put(-203,121){\footnotesize $\mathbf{LN}$}  
  \put(-299,121){\footnotesize $\mathbf{LN}$}  
  \put(-206,138){\footnotesize $\mathbf{FFN}$}  
  \put(-302,138){\footnotesize $\mathbf{FFN}$} 
  \put(-298,168){\footnotesize $\mathbf{LP}$}  
  \put(-202,168){\footnotesize $\mathbf{LP}$} 
  \put(-302,62){\footnotesize $\mathbf z^{lp}_{tar}$} 
  \put(-300,16){\footnotesize $\mathbf z^{0}_{tar}$} 
  \put(-278,62){\footnotesize $\mathbf z^{0}_{aux}$} 
  \put(-230,62){\footnotesize $\mathbf z^{0}_{tar}$} 
  \put(-205,16){\footnotesize $\mathbf z^{0}_{aux}$} 
  \put(-204,62){\footnotesize $\mathbf z^{lp}_{aux}$} 
  \put(-287,108){\footnotesize $\mathbf z^{CA}_{tar}$}  
  \put(-220,108){\footnotesize $\mathbf z^{CA}_{aux}$}  
  \put(-302,203){\footnotesize $\mathbf z^{N}_{tar}$} 
  \put(-202,203){\footnotesize $\mathbf z^{N}_{aux}$} 
  \put(-141,49){\footnotesize $\mathbf Q$}
  \put(-118,49){\footnotesize $\mathbf K$}
  \put(-97,49){\footnotesize $\mathbf V$}
  \put(-61,49){\footnotesize $\mathbf V$}
  \put(-40,49){\footnotesize $\mathbf K$}
  \put(-17,49){\footnotesize $\mathbf Q$}
  \put(-142,73){$\operatorname {Softmax}$}
  \put(-42,73){$\operatorname {Softmax}$}  
  \put(-121,109){\footnotesize $\mathbf{LP}$}  
  \put(-42,109){\footnotesize $\mathbf{LP}$}  
  \put(-129,34){\small $\operatorname {Concate}$}  
  \put(-59,34){\small $\operatorname {Concate}$}   
  \put(-107,93){\small $\mathbf z^{sa}_{tar}$}  
  \put(-65,93){\small $\mathbf z^{sa}_{aux}$}  
  \put(-142,18){\small ${}_{tar}$}  
  \put(-90,18){\small ${}_{aux}$}    
  \put(-70,18){\small ${}_{tar}$}  
  \put(-20,18){\small ${}_{aux}$}  
  \put(-124,122){\small $\mathbf z^{CA}_{tar}$}  
  \put(-45,122){\small $\mathbf z^{CA}_{aux}$} 
  \caption{(a) Architecture of the \textbf{multi-modal transformer} for multi-modal feature fusion, which is a cascade of several \textbf{cross transformer encoder} (CT Encoder) modules. {\color{green} Green} arrows correspond to information updates while {\color{red}red} arrows facilitate information exchange between the two modalities. (b) An illustration of the \textbf{cross transformer encoder}. Our cross transformer encoder consists of cross attention$_{tar}$ and cross attention$_{aux}$ modules with different patch sizes, enabling it to produce stronger features for the restoration of the target modality. The {\color{red}red} arrows facilitate information exchange between the two modalities. (c) \textbf{Cross attention module}. The features of the auxiliary branch are aligned with the target branch of large patch size, while the features of the target branch are aligned with the auxiliary branch of large patch size.}
  \label{figabc} 
\end{figure*}
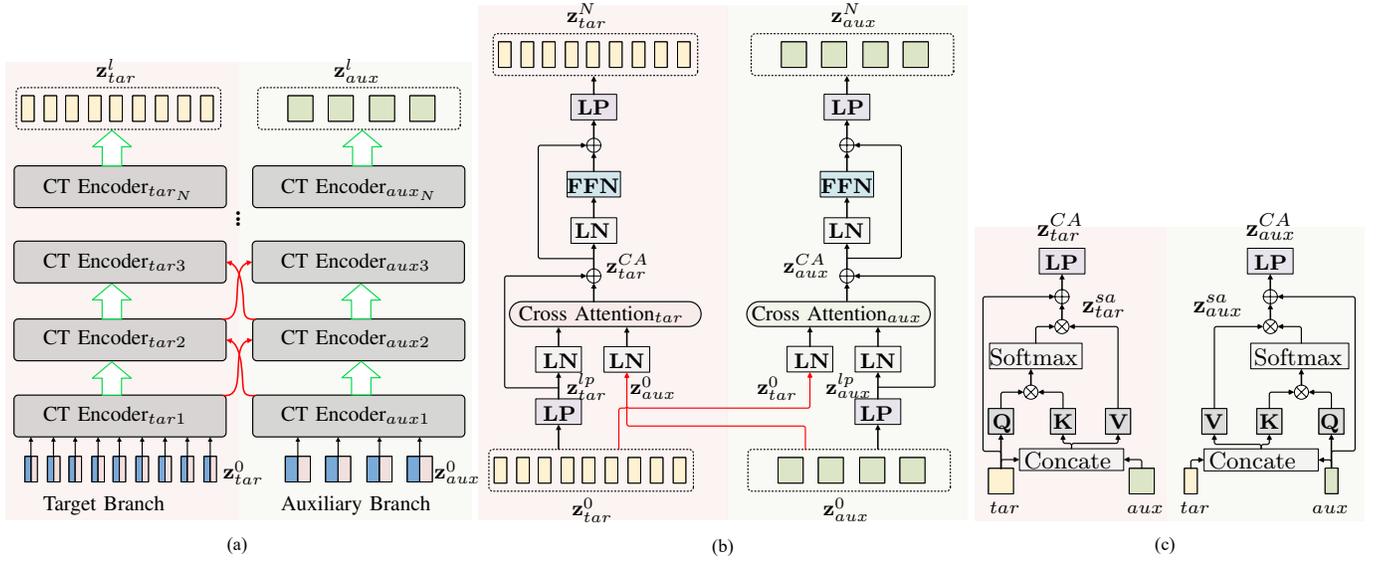
Our multi-modal transformer fuses the different modalities, as shown in~\figref{figabc} (a), which employs two symmetric branches, \eg, a target branch and an auxiliary branch. 
To handle 2D images, following~\cite{dosovitskiy2020image}, we split the features of the two modalities $\mathbf{F}_{aux}$ and $\mathbf{F}_{tar}$ into patches, which are regarded as sequences of "words". Specifically, we first reshape the features of the auxiliary modality $\mathbf{F}_{aux} \in \mathbb{R}^{C \times H \times W}$ into a sequence of patches $\mathbf{F}^{p}_{aux_{i}} \in \mathbb{R}^{P^2 \times C}, i=\{1, \ldots, N\}$, where $N=\frac{H W}{P^{2}}$ is the number of patches or the length of the sequence for the transformer, and $P$ is the resolution of each image patch. Similarly, we reshape the features of the target modality $\mathbf{F}_{tar}$ into patches $\mathbf{F}^{p}_{tar_{i}} \in \mathbb{R}^{(\frac{P}{2})^2 \times C}, i=\{1, \ldots, N\}$. Note that the resolution of the target image patch is a quarter the size of the auxiliary modality. We use the different-sized image patches in the auxiliary and target modalities to produce stronger image features. Then, learnable position encodings $\mathbf E^{p}_{{tar_{i}}} \in \mathbb{R}^{(\frac{P}{2})^2 \times C}$ for the target branch and $\mathbf E^{p}_{{aux_{i}}} \in \mathbb{R}^{P^{2} \times C}$ for the auxiliary modality are added to each branch to maintain the position information of each image patch~\cite{dosovitskiy2020image,carion2020end}: 
\begin{equation}
\footnotesize
\begin{aligned}
\mathbf z^{0}_{tar}&=\left[\mathbf E^{p}_{{tar_{1}}} + \mathbf{F}^{p}_{tar_{1}}, \mathbf E^{p}_{{tar_{2}}} + \mathbf{F}^{p}_{tar_{2}}, \ldots, \mathbf E^{p}_{{tar_{i}}} + \mathbf{F}^{p}_{tar_{i}}\right],\\
\mathbf z^{0}_{aux}&=\left[\mathbf E^{p}_{{aux_{1}}} + \mathbf{F}^{p}_{aux_{1}}, \mathbf E^{p}_{{aux_{2}}} + \mathbf{F}^{p}_{aux_{2}}, \ldots, \mathbf E^{p}_{{aux_{i}}} + \mathbf{F}^{p}_{aux_{i}}\right],
\end{aligned}
\end{equation}
where $\mathbf z^{0}_{tar}\in \mathbb{R}^{(\frac{P}{2})^2 \times C}$ and $\mathbf z^{0}_{aux}\in \mathbb{R}^{P^{2} \times C}$ are the position-embedded patches of the target and auxiliary modality, which are sent to a series of cascaded cross transformer encoder modules (see \figref{figabc} (a)). Each cross transformer encoder consists of two components, \eg, a cross transformer encoder$_{tar}$ for the target modality and cross transformer encoder$_{aux}$ for the auxiliary modality. Note that encoder$_{tar}$ fuses the features from the auxiliary modality, while encoder$_{aux}$ fuses the features from the target modality. Such a cross pattern ensures that each branch learns important information from the other modality. The {\color{green}green} arrows in~\figref{figabc} (a) correspond to information updating for the modality of the current branch, and the {\color{red}red} arrows facilitate information exchange between the two modalities. We can formulate our multi-modal transformer as:
\begin{equation}\label{eq:3}
    [\mathbf z^{N}_{tar},\mathbf z^{N}_{aux}] =\mathcal{M}^N([\mathbf z^{0}_{tar},\mathbf z^{0}_{aux}]),
\end{equation}
where $\mathcal{M}^N$ is the multi-modal transformer module consisting of the $N$-th cross transformer encoder, and $\mathbf z^{N}_{aux}\in \mathbb{R}^{P^{2} \times C}$ and $\mathbf z^{N}_{tar}\in \mathbb{R}^{(\frac{P}{2})^2 \times C}$ are the output sequences of the multi-modal transformer.

\subsubsection{Cross Transformer Encoder}


Our cross transformer encoder aims to effectively fuse the two modalities. As shown in~\figref{figabc} (b), the position-embedded patches $\mathbf z^{0}_{tar}$, $\mathbf z^{0}_{aux}$ are first linearly projected ($\operatorname {LP}$) to align their dimensions, which can be formulated as:
\begin{equation}
\begin{aligned}
   \mathbf z^{lp}_{tar} &= \operatorname {LP}(\mathbf z^{0}_{tar}), \mathbf z^{lp}_{tar} \in \mathbb{R}^{P^{2} \times C},\\
   \mathbf z^{lp}_{aux} &= \operatorname {LP}(\mathbf z^{0}_{aux}), \mathbf z^{lp}_{aux} \in \mathbb{R}^{(\frac{P}{2})^2 \times C},
\end{aligned}
\end{equation}
where $\mathbf z^{lp}_{tar}$ and $\mathbf z^{lp}_{aux}$ are the aligned features. 
These features are sent to the cross attention module and a layernorm to fuse the two modalities. Then, followed by~\cite{vaswani2017attention}, we use a feed-forward network ($\operatorname{FFN}$), which consists of two linear transformations with a ReLU activation in between, to project the feature obtained by the cross attention to a larger space and extract the required information. This can be formulated as:

\begin{equation}
\begin{aligned}
\mathbf z^{CA}_{tar} &= \operatorname{CA_{tar}}\left(\operatorname{LN}\left(\mathbf z^{lp}_{tar}, \mathbf z^{0}_{aux} \right)\right) +  \mathbf z^{lp}_{tar}, \\
\mathbf z^{CA}_{aux} &= \operatorname{CA_{aux}}\left(\operatorname{LN}\left(\mathbf z^{0}_{tar}, \mathbf z^{lp}_{aux} \right)\right) +  \mathbf z^{lp}_{aux},\\
\mathbf z^{i}_{tar} &= \operatorname {LP}\left(\operatorname{FFN}\left(\operatorname{LN}\left(\mathbf z^{CA}_{tar}\right)\right)+\mathbf z^{CA}_{tar}\right),\\
\mathbf z^{i}_{aux} &= \operatorname {LP}\left( \operatorname{FFN}\left(\operatorname{LN}\left(\mathbf z^{CA}_{aux}\right)\right)+\mathbf z^{CA}_{aux}\right),\\
\end{aligned}
\end{equation}
where $i = [1,2,...,N]$, $\operatorname{CA_{tar}}$ and $\operatorname{CA_{aux}}$ are the cross attention modules for different modalities. $\operatorname{LN}$ denotes the layer normalization that aims to normalize the hidden layer in the neural network to the standard normal distribution to accelerate training speed and convergence~\cite{vaswani2017attention}. The output sequence features $\mathbf z^{i}_{aux}$ and $\mathbf z^{i}_{tar}$ of the two branches are saved as the input of the next cross transformer encoder.

\subsubsection{Cross Attention Module}
Our cross attention module is an improved multi-head attention module which absorbs features from the auxiliary modality that contribute to the target modality. Specifically, in order to fuse the different modalities more efficiently and effectively, the features in the current branch serve as a query that interacts with the features from the other branch through attention. Note that the query features have already been dimensionally aligned with the features from the other branch. In other words, the feature sizes in the two branches are different. This allows our cross attention fusion module to learn both high-level context features and local details.
An illustration of our symmetric cross attention module is shown in~\figref{figabc} (c). For the target branch, we use the aligned features $\mathbf z^{lp}_{tar}$ after layer normalization $\operatorname{LN}(\mathbf z^{lp}_{tar})\in \mathbb{R}^{P^{2} \times C}$ as the query ($\mathbf{Q}$), and concatenate them with the features from the auxiliary branch $\operatorname{LN}(\mathbf z^{0}_{aux})\in \mathbb{R}^{P^{2} \times C}$ to serve as the key ($\mathbf{K}$) and value ($\mathbf{V}$). Similarly, for the auxiliary branch, the aligned features $\mathbf z^{lp}_{aux}$ after layer normalization $\operatorname{LN}(\mathbf z^{lp}_{aux})\in \mathbb{R}^{(\frac{P}{2})^2 \times C}$ serve as $\mathbf{Q}$, and these are concatenated with the features from the target branch $\operatorname{LN}(\mathbf z^{0}_{tar})\in \mathbb{R}^{(\frac{P}{2})^2 \times C}$ to serve as $\mathbf{K}$ and $\mathbf{V}$. Then, the correspondence between the two modalities in each branch can be found using the following bilinear model:
\begin{equation}
\mathbf z^{sa}_{} = \operatorname{softmax}\left(\frac{\mathbf Q \mathbf K^{T}}{\sqrt{\mathbf D / h}}\right) \mathbf V,
\end{equation}
where $\mathbf z^{sa}_{}$ can be expressed as $\mathbf z^{sa}_{tar}$ for the target branch, and $\mathbf z^{sa}_{aux}$ for the auxiliary branch, $\mathbf D$ is the embedding dimension of $\operatorname{LN}$ and $h$ is the number of heads in the cross attention mechanism. Finally, the outputs $\mathbf z^{CA}_{tar}$ and $\mathbf z^{CA}_{aux}$ of a cross attention module can be defined as follows:
\begin{equation}
\begin{aligned}
\mathbf z^{CA}_{tar} &= \operatorname{LP}(\mathbf z^{sa}_{tar} + \operatorname{LN}(\mathbf z^{lp}_{tar})),\\
\mathbf z^{CA}_{aux} &= \operatorname{LP}(\mathbf z^{sa}_{aux} + \operatorname{LN}(\mathbf z^{lp}_{aux})).
\end{aligned}
\end{equation}

\begin{figure}[t]
\centering
  \includegraphics[width=0.5\textwidth]{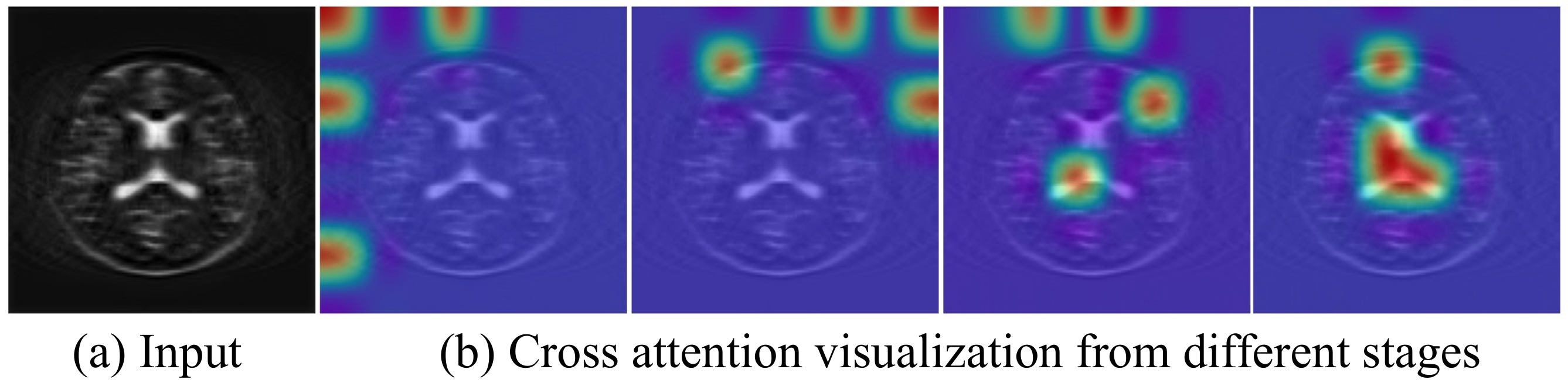}
  \caption{\textbf{Cross attention visualization from different stages}, where the four bright spots indicate the areas that four heads focus on. The network gradually focuses on the brain regions as the increase of network depth.}
  \label{attentionmap} 
\end{figure}

To qualitatively evaluate the mechanism of our cross attention module, we visualize the cross attention map of four cascaded cross transformer encoders in~\figref{attentionmap}, in which the four bright spots indicate the areas focused by the four heads~\cite{sun2021getam}. As can be observed, with the number of network layers deepens, the attention of the network gradually moves towards the middle (object) region. Specifically, attention maps from different stages focus on different regions of the image (\eg, the shallow stages mainly focus on the background area, and the deep stages mainly focuses on the object area). Obviously, the information in the background area is useless for the MR image reconstruction. The goal of the cross-attention module is to effectively integrate two modalities, thereby the clear structural information can be learned from the auxiliary branch to assist the reconstruction of target modality. The attention of the network in this figure is gradually moving towards the object region, which precisely indicates that the cross attention mechanism can push the network to learn clear structural details from the object region in the auxiliary modality.

\subsection{Tails}
Finally, the outputs of the multi-modal transformer encoder $\mathbf z^{N}_{tar}$ and $\mathbf z^{N}_{aux}$ are fed into two tail modules to predict the restored images, each of which consists of a 1$\times$1 convolutional layers:
\begin{equation}\label{eq:excite} 
\mathbf x'_{tar} = \operatorname{Tail}_{tar}\left(\mathbf z^{N}_{tar}\right), \; \text{and} \;
\mathbf x'_{aux} = \operatorname{Tail}_{aux}\left(\mathbf z^{N}_{aux}\right).
\end{equation}
For the reconstruction task, each tail module consists of three convolutional layers. Please note, for the SR task, we add a sub-pixel convolutional layer~\cite{zhang2018residual} to $\operatorname{Tail}_{tar} $ to upscale the output $\mathbf x'_{tar}$ is an image of size $H \times W$.

\subsection{Loss}
Following~\cite{sriram2020grappanet,wang2020deepcomplexmri}, we simply use the $L_1$ loss to evaluate
our results:
\begin{equation}
\small
L=\frac{1}{M} \sum_{m=1}^{M} \alpha\left\|{{\mathbf x}'}_{tar}^{m}-{{\mathbf x}}_{tar}^{m}\right\|_{1}+(1-\alpha)\left\|{{\mathbf x}'}_{aux}^{m}-{{\mathbf x}}_{aux}^{m}\right\|_{1},
\end{equation}
where $\alpha$ weights the trade-off between the two modalities, and $M$ is the number of training samples. Note that the loss on the auxiliary modality is to encode the auxiliary modality and ensure that the features of the auxiliary modality can be fully extracted.

\section{Experiments}\label{sec:experiment}

\begin{figure*}[!t]

	\makeatletter\def\@captype{table}\makeatother\caption{Average (with standard deviation) \textbf{reconstruction} results, in terms of SSIM, PSNR and NMSE, under different undersampling patterns. The best and second-best results are marked in {\color{red}red} and {\color{blue}blue}, respectively. $P<$ 0.001 was considered as a statistically significant level.}  
    \centering
	\begin{minipage}[c]{0.48\textwidth}
		\centering
        \label{table1}
		\begin{threeparttable}
			\resizebox{1\textwidth}{!}{
				\setlength\tabcolsep{4pt}
				\renewcommand\arraystretch{1.1}
				\begin{tabular}{r||cccc}
			        \hline\thickhline
			        \rowcolor{mygray}
			        {\textbf{fastMRI}}&  \multicolumn{4}{c}{Random 4$\times$} \\\hline
			        \rowcolor{mygray}

			     {Method}
			        &~~SSIM$\uparrow$~~ &~~PSNR$\uparrow$~~ &~~NMSE$\downarrow$~~& $P$ values\\ \hline\hline
                    Zero-filling &0.442$\pm$0.10 &24.5$\pm$1.37 &0.057$\pm$0.05&$<0.001$/$<0.001$/$<0.001$\\ 
                    LORAKS~\cite{haldar2013low}  &0.530$\pm$0.07 &26.0$\pm$1.03 &0.050$\pm$0.02 &$<0.001$/$<0.001$/$<0.001$\\
                    MoDL~\cite{aggarwal2018modl} &0.576$\pm$0.04 &27.7$\pm$0.96 &0.047$\pm$0.01 &$<0.001$/$<0.001$/$<0.001$\\
                    TransMRI~\cite{vaswani2017attention}  &0.607$\pm$0.05 &28.4$\pm$0.81 &0.038$\pm$0.01&$<0.001$/$=0.001$/$<0.001$ \\ \hline
                    Transmed~\cite{dai2021transmed} &0.609$\pm$0.06 &28.4$\pm$0.99 &0.040$\pm$0.03 &$<0.001$/$<0.001$/$<0.001$\\
                    HyperDense-Net~\cite{dolz2018hyperdense} &0.600$\pm$0.04 &28.3$\pm$1.00 &0.038$\pm$0.02 &$<0.001$/$<0.001$/$<0.001$\\
                    MDUNet~\cite{xiang2018deep} &0.600$\pm$0.05 &28.6$\pm$1.00 &0.040$\pm$0.01&$<0.001$/$<0.001$/$<0.001$\\
                    rsGAN~\cite{dar2020prior} &{\color{blue}0.608$\pm$0.04} &{\color{blue}28.9$\pm$1.03} &{\color{blue}0.033$\pm$0.02}&$<0.001$/$<0.001$/$<0.001$\\ 
                    \textbf{MTrans}   &{\color{red}0.638$\pm$0.03} &{\color{red}29.3$\pm$0.89} &{\color{red}0.030$\pm$0.01}&$-$\\ 
           
                    \hline
	        	\end{tabular}
			}

		\end{threeparttable}
		\vspace*{2pt}
	\end{minipage} 
	\begin{minipage}[c]{0.48\textwidth}
		\centering
		\begin{threeparttable}
			\resizebox{1\textwidth}{!}{
				\setlength\tabcolsep{4pt}
				\renewcommand\arraystretch{1.1}
				\begin{tabular}{r||cccc}
			        \hline\thickhline
			        \rowcolor{mygray}
			        {\textbf{uiMRI}}&  \multicolumn{4}{c}{Random 6$\times$} \\\hline
			        \rowcolor{mygray}

			     {Method}
			        &~~SSIM$\uparrow$~~ &~~PSNR$\uparrow$~~ &~~NMSE$\downarrow$~~& $P$ values\\ \hline\hline
                    Zero-filling &0.700$\pm$0.09 &27.0$\pm$1.70 &0.067$\pm$0.010&$<0.001$/$<0.001$/$<0.001$\\
                    LORAKS~\cite{haldar2013low}  &0.746$\pm$0.07 &27.5$\pm$1.27 &0.057$\pm$0.005 &$<0.001$/$<0.001$/$<0.001$\\
                    MoDL~\cite{aggarwal2018modl} &0.826$\pm$0.04 &28.4$\pm$1.16 &0.048$\pm$0.007 &$<0.001$/$<0.001$/$<0.001$\\
                    TransMRI~\cite{vaswani2017attention} &0.861$\pm$0.02 &28.9$\pm$1.32 &0.044$\pm$0.005 &$<0.001$/$<0.001$/$<0.001$\\ \hline
                    Transmed~\cite{dai2021transmed} &0.879$\pm$0.04 &29.3$\pm$0.99 &0.040$\pm$0.004 &$<0.001$/$=0.002$/$<0.001$\\
                    HyperDense-Net~\cite{dolz2018hyperdense} &0.880$\pm$0.04 &29.7$\pm$1.11 &0.038$\pm$0.006 &$<0.001$/$=0.002$/$<0.001$\\
                    MDUNet~\cite{xiang2018deep} &0.900$\pm$0.03 &30.0$\pm$1.50 &0.034$\pm$0.006 &$<0.001$/$<0.001$/$<0.001$\\ 
                    rsGAN~\cite{dar2020prior} &{\color{blue}0.908$\pm$0.03} &{\color{blue}30.7$\pm$1.42} &{\color{blue}0.028$\pm$0.005}&$<0.001$/$<0.001$/$<0.001$ \\ 
                    \textbf{MTrans}   &{\color{red}0.931$\pm$0.02} &{\color{red}31.7$\pm$1.13} &{\color{red}0.024$\pm$0.004} &$-$\\ \hline
	        	\end{tabular}
			}
		\end{threeparttable}
		\vspace*{2pt}
	\end{minipage}

	\begin{minipage}[c]{0.48\textwidth}
		\centering
        \label{table1}
		\begin{threeparttable}
			\resizebox{1\textwidth}{!}{
				\setlength\tabcolsep{4pt}
				\renewcommand\arraystretch{1.1}
				\begin{tabular}{r||cccc}
			        \hline\thickhline
			        \rowcolor{mygray}
			        {\textbf{fastMRI}}&  \multicolumn{4}{c}{Equispaced 8$\times$} \\\hline
			        \rowcolor{mygray}

			     {Method}
			        &~~SSIM$\uparrow$~~ &~~PSNR$\uparrow$~~ &~~NMSE$\downarrow$~~& $P$ values\\ \hline\hline
                    Zero-filling &0.369$\pm$0.12 &22.9$\pm$1.25 &0.094$\pm$0.04&$<0.001$/$<0.001$/$<0.001$\\ 
                    LORAKS~\cite{haldar2013low}  &0.443$\pm$0.05 &24.7$\pm$1.05 &0.099$\pm$0.05 &$<0.001$/$<0.001$/$<0.001$\\
                    MoDL~\cite{aggarwal2018modl} &0.437$\pm$0.04 &24.4$\pm$0.86 &0.083$\pm$0.04 &$<0.001$/$<0.001$/$<0.001$\\
                    UNet~\cite{zbontar2018fastmri} &0.498$\pm$0.05 &26.3$\pm$1.00 &0.054$\pm$0.03&$<0.001$/$<0.001$/$<0.001$\\ 
                    TransMRI~\cite{vaswani2017attention}  &0.507$\pm$0.05 &26.8$\pm$0.83 &0.048$\pm$0.02&$<0.001$/$<0.001$/$<0.001$ \\ \hline
                    Transmed~\cite{dai2021transmed} &0.518$\pm$0.06 &27.1$\pm$0.78 &0.057$\pm$0.03 &$<0.001$/$=0.02$/$<0.001$\\
                    HyperDense-Net~\cite{dolz2018hyperdense} &0.521$\pm$0.04 &27.3$\pm$0.90 &{\color{blue}0.046$\pm$0.03} &$<0.001$/$<0.001$/$<0.001$\\
                    MDUNet~\cite{xiang2018deep} &{\color{blue}0.544$\pm$0.05} &{\color{blue}27.9$\pm$0.86} &{\color{blue}0.046$\pm$0.02}&$<0.001$/$<0.001$/$<0.001$\\ 
                    rsGAN~\cite{dar2020prior} &0.530$\pm$0.04 &27.6$\pm$0.80 &0.049$\pm$0.02&$<0.001$/$<0.001$/$<0.001$\\ 
                    \textbf{MTrans}   &{\color{red}0.563$\pm$0.04} &{\color{red}28.4$\pm$0.81} &{\color{red}0.043$\pm$0.02}&$-$\\ \hline
	        	\end{tabular}
			}

		\end{threeparttable}
		\vspace*{2pt}
	\end{minipage} 
	\begin{minipage}[c]{0.48\textwidth}
		\centering
		\begin{threeparttable}
			\resizebox{1\textwidth}{!}{
				\setlength\tabcolsep{4pt}
				\renewcommand\arraystretch{1.1}
				\begin{tabular}{r||cccc}
			        \hline\thickhline
			        \rowcolor{mygray}
			        {\textbf{uiMRI}}&  \multicolumn{4}{c}{Equispaced 8$\times$} \\\hline
			        \rowcolor{mygray}

			     {Method}
			        &~~SSIM$\uparrow$~~ &~~PSNR$\uparrow$~~ &~~NMSE$\downarrow$~~& $P$ values\\ \hline\hline
                    Zero-filling &0.560$\pm$0.10 &24.8$\pm$1.76 &0.087$\pm$0.010&$<0.001$/$<0.001$/$<0.001$\\
                    LORAKS~\cite{haldar2013low}  &0.699$\pm$0.06 &26.0$\pm$1.43 &0.061$\pm$0.004 &$<0.001$/$<0.001$/$<0.001$\\
                    MoDL~\cite{aggarwal2018modl} &0.726$\pm$0.04 &27.3$\pm$1.04 &0.058$\pm$0.002 &$<0.001$/$<0.001$/$<0.001$\\
                    UNet~\cite{zbontar2018fastmri} &0.713$\pm$0.06 &27.0$\pm$1.39 &0.055$\pm$0.006&$<0.001$/$<0.001$/$<0.001$\\ 
                    TransMRI~\cite{vaswani2017attention} &0.721$\pm$0.04 &27.4$\pm$1.11 &0.054$\pm$0.007&$<0.001$/$<0.001$/$<0.001$ \\ \hline
                    Transmed~\cite{dai2021transmed} &0.780$\pm$0.06 &28.0$\pm$1.20 &0.052$\pm$0.005 &$<0.001$/$<0.001$/$<0.001$\\
                    HyperDense-Net~\cite{dolz2018hyperdense} &0.794$\pm$0.05 &28.1$\pm$1.01 &0.049$\pm$0.002 &$<0.001$/$<0.001$/$<0.001$\\
                    MDUNet~\cite{xiang2018deep} &0.820$\pm$0.03 &28.3$\pm$1.07 &0.048$\pm$0.005 &$<0.001$/$<0.001$/$<0.001$\\ 
                    rsGAN~\cite{dar2020prior} &{\color{blue}0.878$\pm$0.03} &{\color{blue}28.7$\pm$1.35} &{\color{blue}0.040$\pm$0.006} &$<0.001$/$<0.001$/$<0.001$\\ 
                    \textbf{MTrans}   &{\color{red}0.910$\pm$0.02} &{\color{red}30.8$\pm$1.07} &{\color{red}0.032$\pm$0.003}&$-$ \\ \hline
	        	\end{tabular}
			}
		\end{threeparttable}
		\vspace*{2pt}
	\end{minipage}

	\hfill
\end{figure*}

\begin{figure*}[!t]

	\makeatletter\def\@captype{table}\makeatother\caption{Average (with standard deviation) \textbf{super-resolution} results, in terms of SSIM, PSNR and NMSE, under different datasets. The best and second-best results are marked in {\color{red}red} and {\color{blue}blue}, respectively. $P<$ 0.001 was considered as a statistically significant level.} 
	\centering
	\begin{minipage}[t]{0.48\textwidth}
		\centering
        \label{table2}
		\begin{threeparttable}
			\resizebox{1\textwidth}{!}{
				\setlength\tabcolsep{4pt}
				\renewcommand\arraystretch{1.1}
				\begin{tabular}{r||cccc}
			        \hline\thickhline
			        \rowcolor{mygray}
			        {\textbf{fastMRI}}&  \multicolumn{4}{c}{4$\times$} \\\hline
			        \rowcolor{mygray}

			     {Method}
			        &~~SSIM$\uparrow$~~ &~~PSNR$\uparrow$~~ &~~NMSE$\downarrow$~~& $P$ values \\ \hline\hline
                    Bicubic &0.400$\pm$0.07 &16.9$\pm$1.70 &0.917$\pm$0.06&$<0.001$/$<0.001$/$<0.001$\\ 
                    EDSR~\cite{lim2017enhanced} &0.580$\pm$0.04 &28.1$\pm$1.64 &0.045$\pm$0.04&$<0.001$/$<0.001$/$<0.001$\\ 
                    TransMRI~\cite{vaswani2017attention} &0.600$\pm$0.03 &29.9$\pm$1.44 &0.048$\pm$0.02&$<0.001$/$<0.001$/$<0.001$\\ \hline
                    Transmed~\cite{dai2021transmed} &0.673$\pm$0.05 &29.4$\pm$1.35 &0.053$\pm$0.05 &$<0.001$/$<0.001$/$<0.001$\\
                    HyperDense-Net~\cite{dolz2018hyperdense} &0.640$\pm$0.04 &30.4$\pm$1.62 &0.042$\pm$0.04 &$<0.001$/$<0.001$/$<0.001$\\
                    PRO~\cite{lyu2020multi} &0.700$\pm$0.02 &30.8$\pm$1.60 &0.038$\pm$0.03&$<0.001$/$<0.001$/$<0.001$\\
                    MCSR~\cite{zeng2018simultaneous} &{\color{blue}0.704$\pm$0.03} &{\color{blue}31.0$\pm$1.31} &{\color{blue}0.033$\pm$0.03}&$<0.001$/$<0.001$/$<0.001$\\
                    \textbf{MTrans}   &{\color{red}0.719$\pm$0.02} &{\color{red}31.9$\pm$1.19} &{\color{red}0.031$\pm$0.02}&$-$ \\ \hline
	        	\end{tabular}
			}

		\end{threeparttable}
		\vspace*{2pt}
	\end{minipage}
	\begin{minipage}[t]{0.48\textwidth}
		\centering
		\begin{threeparttable}
			\resizebox{1\textwidth}{!}{
				\setlength\tabcolsep{4pt}
				\renewcommand\arraystretch{1.1}
				\begin{tabular}{r||cccc}
			        \hline\thickhline
			        \rowcolor{mygray}
			        {\textbf{uiMRI}}&  \multicolumn{4}{c}{4$\times$} \\\hline
			        \rowcolor{mygray}

			     {Method}
			        &~~SSIM$\uparrow$~~ &~~PSNR$\uparrow$~~ &~~NMSE$\downarrow$~~& $P$ values \\ \hline\hline
                    Bicubic &0.526$\pm$0.05 &8.3$\pm$1.20 &0.900$\pm$0.030&$<0.001$/$<0.001$/$<0.001$\\ 
                    EDSR~\cite{lim2017enhanced} &0.941$\pm$0.07 &32.3$\pm$1.04 &0.012$\pm$0.004&$<0.001$/$<0.001$/$<0.001$\\ 
                    TransMRI~\cite{vaswani2017attention}  &0.940$\pm$0.05 &33.5$\pm$1.17 &0.009$\pm$0.005&$<0.001$/$<0.001$/$<0.001$ \\ \hline
                    Transmed~\cite{dai2021transmed} &{\color{blue}0.947$\pm$0.06} &{\color{blue}34.9$\pm$0.78} &{\color{blue}0.006$\pm$0.004} &$<0.001$/$<0.001$/$<0.001$\\
                    HyperDense-Net~\cite{dolz2018hyperdense} &0.940$\pm$0.05 &33.9$\pm$0.90 &0.008$\pm$0.005 &$<0.001$/$<0.001$/$<0.001$\\
                    PRO~\cite{lyu2020multi} &0.945$\pm$0.07 &34.4$\pm$0.97 &0.007$\pm$0.003&$<0.001$/$<0.001$/$<0.001$\\ 
                    MCSR~\cite{zeng2018simultaneous} &0.944$\pm$0.07 &34.8$\pm$0.97 &{\color{blue}0.006$\pm$0.003}&$<0.001$/$<0.001$/$<0.001$\\ 
                    \textbf{MTrans}   &{\color{red}0.959$\pm$0.05} &{\color{red}36.1$\pm$0.79} &{\color{red}0.005$\pm$0.003}&$-$ \\ \hline
	        	\end{tabular}
			}

		\end{threeparttable}
		\vspace*{2pt}
	\end{minipage}

\end{figure*}
In this section, we first introduce the datasets and baselines used in our experiments, followed by the implementation details. Then, we summarize and analyze the experimental results. Finally, we conduct an ablation study to investigate the effectiveness of our multi-modal strategies.

\begin{figure*}[!t]
\centering
  \includegraphics[width=1\textwidth]{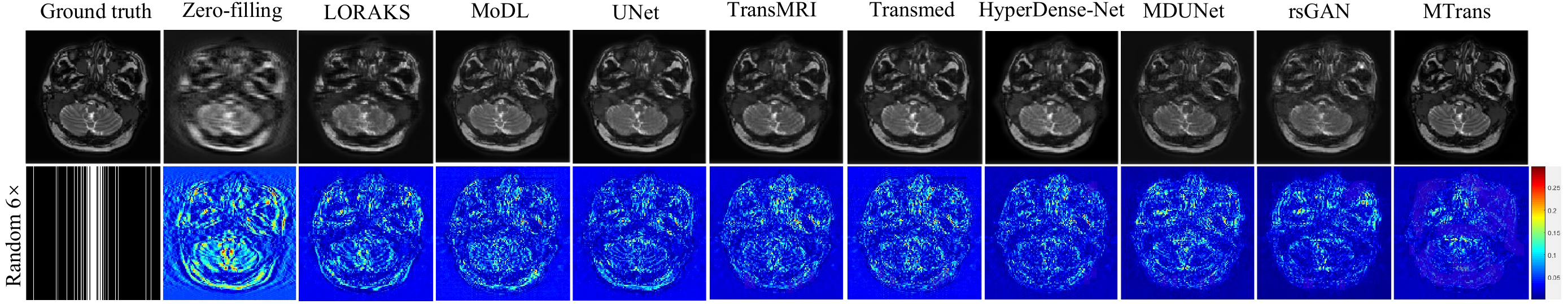}
  \caption{Comparison of different methods in terms of reconstruction results on the uiMRI dataset. Reconstructed images and error maps are presented with corresponding quantitative measurements in PSNR/SSIM. The more obvious errors, the worse the restoration results. The first three methods represent the single-modal results, while the last three represent the multi-modal results.}
  \label{figone} 
\end{figure*}

\begin{figure}[!t]
\centering
  \includegraphics[width=0.47\textwidth]{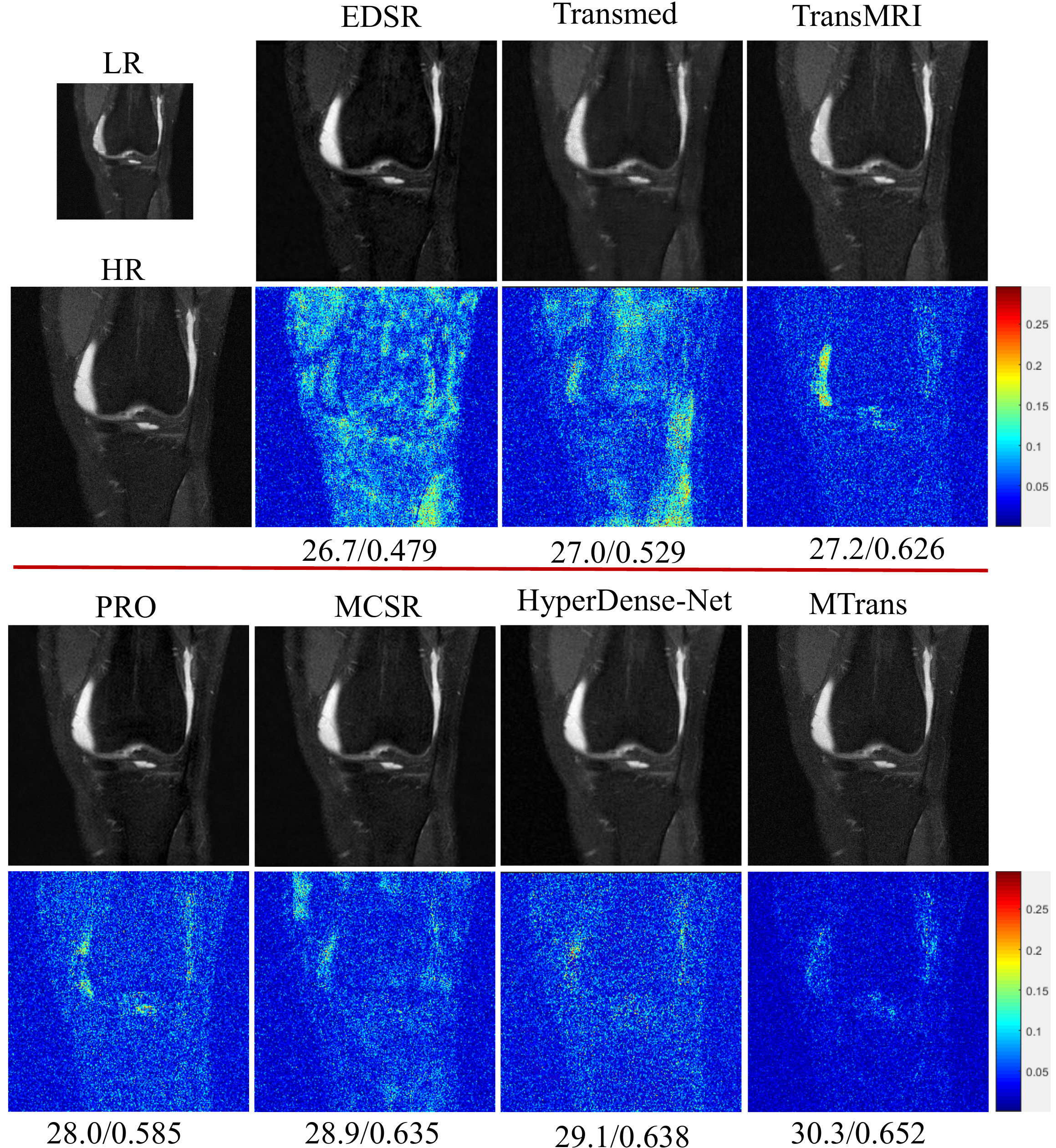}\caption{Comparison of different methods in terms of SR results on the fastMRI dataset. SR images and error maps are presented with corresponding quantitative measurements in PSNR/SSIM. The more obvious the errors, the worse the restoration results. The first two rows show the single-modal results, while the last two rows provide the multi-modal results.}
  \label{figtwo}
\end{figure}

\subsection{Implementation Details}
\subsubsection{Datasets}
We use three raw MR image datasets to evaluate our method: (1) \textbf{fastMRI}~\cite{zbontar2018fastmri} is the largest open-access raw MR image dataset, officially provided at \url{https://fastmri.med.nyu.edu/}. Following~\cite{xuan2020learning}, we filter out 227 and 24 pairs of PDWI and FS-PDWI knee volumes for training and validation. PDWIs are used to guide the restoration of the FS-PDWI modality. (2) \textbf{uiMRI} was collected using a 3T system (provided by United Imaging Healthcare uMR 790; informed written consent was obtained from all the subjects, and all experiments were carried out in accordance with the approved guidelines) with two different protocols (whole brain T1WI and T2WI $k$-space sampling) on 400 subjects. The slice thickness is 4 mm and matrix size is 320$\times$320$\times$19.
The uiMRI dataset is split subject-wise with a ratio of 7:1:2 for the training/validation/test sets, where T1WI is the auxiliary modality and T2WI is the target modality. 
All the experimental datasets are aligned through affine registration before our experiments. (3) \textbf{Multi-coil dataset} was collected using a clinical 3T Siemens Magnetom Skyra scanner, including 20 paired PD and PDFS knee subjects with 15 coils and a matrix size of $320\!\times320\!\times\!20$~\cite{12}. For the fastMRI and uiMRI datasets, the model was trained on the single-coil magnitude images. Note that the undersampled data was obtained on the original $k$-space domain. For the multi-coil dataset, the model was trained on the multi-coil complex images with real and imaginary channel inputs. For the final single-channel reconstruction, the coil sensitivity maps derived from ESPIRiT are used for coil combination~\cite{uecker2014espirit}.

For the reconstruction task, we use the random and Equispaced undersampling patterns with 4$\times$, 6$\times$ and 8$\times$ accelerations. For the SR task, we use 4$\times$ enlargement for both fastMRI and uiMRI to evaluate the effectiveness of our method. The LR image was obtained by truncating the outer part of the fully sampled $k$-space $\mathbf{y}$ with the desired factor to degrade the resolution, and then applying $\mathcal{F}$ to obtain the degraded LR image $\tilde{{\mathbf{x}}}$. This process mimics the real image acquisition process and avoids checkerboard artifacts~\cite{chen2018efficient}.

\subsubsection{Baselines and Training Details}
We compare our model with the following single- and multi-modal algorithms. Hyperparameter optimization for each baseline in our experiments is performed via cross-validation. Single-modal MR image reconstruction/SR methods include: The most popular SR method (EDSR)~\cite{lim2017enhanced}, where Adam is adopted as the optimizer and the learning rate is initialized as $10^{-4}$; The end to end architecture, UNet, provided by fastMRI~\cite{zbontar2018fastmri}, which is retrained with SGD optimizer and a learning rate of 1e-4; A standard transformer framework that contains an encoder-decoder structure with multi-head attention for MR image reconstruction/SR (TransMRI)~\cite{vaswani2017attention}, where the image has been split into patches and treated the same way as tokens (words); MoDL~\cite{aggarwal2018modl}, a model-based unrolled architecture for inverse problems, where the number of layers is set to 5 and number of iterations is set to 10; and LORAKS~\cite{haldar2013low}, a classic compressed-sensing method, where the regularization parameters is set to $10^{-10}$, the matrix rank is set to 50, and the $k$-space neighborhood is set to 5.
Multi-modal reconstruction/SR methods include: A DenseUNet model for multi-modal MR image reconstruction, called MDUNet~\cite{xiang2018deep}, where Adam is adopted as the optimizer and the learning rate is initialized as $10^{-4}$; HyperDense-Net~\cite{dolz2018hyperdense}, a multi-modal segmentation method with feature fusion at intermediate layers; Transmed~\cite{dai2021transmed}, a transformer-based classification method that tries to establish long-range dependencies from the decomposed multi-modal image patches, where SGD is adopted as the optimizer and the learning rate is $10^{-3}$; A conditional GAN framework for multi-modal MR image reconstruction, named rsGAN~\cite{dar2020prior}, where the hyperparameters $\lambda_p$ is set to 100 and $\lambda_perc$ is set to 70; A non-progressive multi-modal MR image SR network called PRO~\cite{lyu2020multi}, where the hyperparameters $\lambda_1$, $\lambda_2$, and $\lambda_3$  in the objective function of the generator are set to $10^{-1}$, $10^{-2}$, and $10^{-11}$; and a deep CNN model for multi-modal MR image SR named MCSR~\cite{zeng2018simultaneous}, where the hyperparameters $\lambda_1$, and $\lambda_2$, are set to 0.6 and 0.4, respectively. For the multi-coil data, we use traditional multi-coil imaging methods (SPIRiT~\cite{lustig2010spirit} and L1-SPIRiT~\cite{murphy2010clinically}, where the the kernel size is set to 5$\times$5.) as well as CNN-based methods (VN-Net~\cite{12} and MoDL~\cite{aggarwal2018modl} as the baselines. For some of the methods in the baselines which have data consistency, we still retain their original structure and ensure that the training is performed under optimal parameters.


Our model is implemented in PyTorch with four NVIDIA Tesla V100 GPUs and 32GB of memory per card. We use the SGD optimizer with a learning rate of 1e-4 and a mini-batch size of 8, and train our model over 50 epochs. The parameter $\alpha$ is set to 0.9, the effectiveness of which is verified in the ablation studies. The number of cross attention heads is set to 4, and the number of channels $C$ in the feature map generated by Head$_{tar}$ and Head$_{aux}$ is set to 16. We use $N$ = 4 cross transformer encoders in our network. For quantitative study, peak signal-to-noise ratio (PSNR), structural similarity index (SSIM) and normalized mean square error (NMSE) are used to evaluate the performance of our method~\cite{zbontar2018fastmri}. The various forms of our multi-modal fusion strategies will be discussed in the ablation studies.

\subsection{Results on MR Image Reconstruction}
\subsubsection{Quantitative Evaluation}

We evaluate our reconstruction results by computing the SSIM, PSNR, and NMSE between the restored image and the fully sampled ground truth image. In Table~{\color{blue}\ref{table1}}, we show the results of our reconstruction over the two raw MR image datasets. The first row provides the single-modal MR image reconstruction methods, while the second row includes the CNN-based and transformer-based multi-modal methods and our multi-modal transformer model. From the results, we find that the PSNR and SSIM values of the single-modal UNet are relatively low, especially on the uiMRI dataset. Similarly, the results of the standard single-modal transformer framework TransMRI are lower than other multi-modal methods. However, with the help of the auxiliary modality, the PSNR, SSIM, and NMSE of Transmed, HyperDense-Net, MDUNet, and rsGAN are improved to a certain degree. Thus, we can conclude that the auxiliary modality is complementary to the target modality. However, these methods do not explore complementary information of different modalities at multiple scales. Our MTrans achieves 29.3 dB and 28.4 dB in PSNR on the fastMRI dataset, 31.7 dB, and 30.8 dB on uiMRI.

\subsubsection{Qualitative Evaluation}
For qualitative analysis, we provide the reconstruction results on the uiMRI dataset in~\figref{figone}. The more obvious the structure in the blue error map, the worse the restoration. As can be seen, reconstructions with zero-filling produce significant aliasing artifacts and lose anatomical details. The first two rows in~\figref{figone} show the reconstruction results of single-modal methods. Compared with zero-filling, single-modal methods can somewhat improve the reconstruction. However, multi-modal methods (the last two rows in~\figref{figone}) provide even further improvements, as verified by their corresponding error maps. Notably, our method yields the lowest reconstruction error, better preserving important anatomical details.

\subsection{Results on MR image Super-Resolution}
\subsubsection{Quantitative Evaluation}
We evaluate our MTrans with the competing baseline methods on the SR task in Table~{\color{blue}\ref{table2}}. This table summarizes the 4$\times$ enlargement results of all methods on the two raw MR image datasets. Similar to Table~{\color{blue}\ref{table1}}, the first row shows the single-modal MR image SR methods, while the second row includes the CNN-based multi-modal methods as well as our multi-modal transformer model. As can be seen from the table, the SR results are similar to those obtained for reconstruction. Specifically, the single-modal architectures, whether based on CNNs or transformers, are not as effective as the multi-modal methods. However, the multi-modal CNN methods are not as effective as our model. For example, our method improves the PSNR result of the best multi-modal CNN method from 31.0 dB to 31.9 dB on the fastMRI dataset, while on the uiMRI dataset it increases it from 34.8 dB to 36.1 dB. These results further confirm the effectiveness of our multi-modal approach.

\subsubsection{Qualitative Evaluation}
~\figref{figtwo} shows the 4$\times$ enlargement of target modality images from fastMRI. The first two rows show the images restored by the single-modal methods, while the last two provide show the results of the multi-modal methods. From this figure, we can see that the basic structure of the image can be restored by single-modal methods (\eg, EDSR and TransMRI). However, the methods based on multi-modal fusion significantly improve the results, with smaller errors and clearer structures. In particular, our MTrans produces high-quality images with clear details, minimal checkerboard effects, and less structural loss. Further, our method can effectively restore the entire structure of the knee. This superior performance is attributed to the fact that our MTrans can effectively aggregate MR image information from different modalities to obtain stronger features.

\subsection{Results on Multi-coil Data}
\begin{figure}[!t]
\centering
  \includegraphics[width=0.49\textwidth]{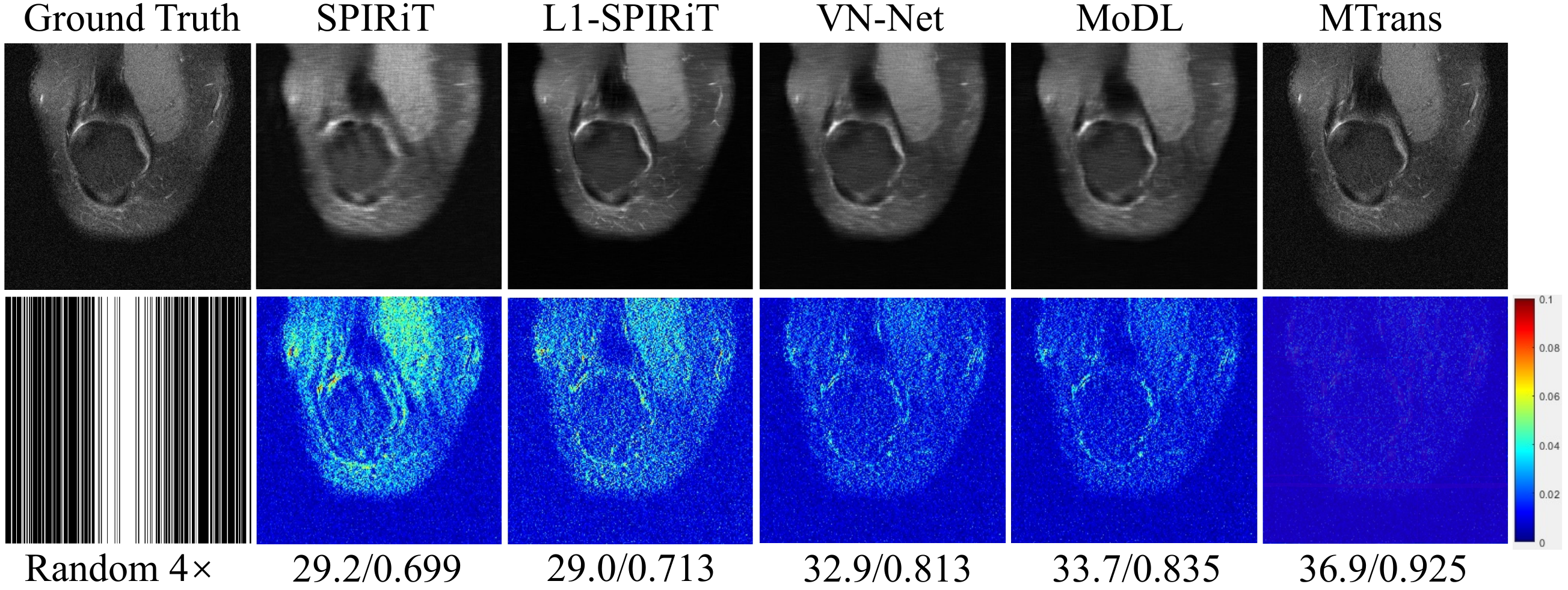}
  \caption{Comparison of different multi-coil reconstruction methods. Reconstructed images and error maps are presented with corresponding quantitative measurements in PSNR/SSIM. The more obvious the errors, the worse the restoration results.}
  \label{figmu}
\end{figure}

To verify the effectiveness of our method on the multi-coil data, we compare our method with various multi-coil reconstruction methods in Table~{\color{blue}\ref{multi-coil}}. As can be seen from this table, the results of model-based methods, such as VN-NET and MoDL, are better than that of traditional methods SPIRiT and L1-SPIRiT. However, benefiting from powerful multi-modal information, our method achieves the best PSNR and SSIM results. We also show the reconstruction images and corresponding error maps in~\figref{figmu}. As can be seen from this figure, our method achieves the lowest texture error and provides results that are very close to the ground-truth. This proves that our method is also effective in multi-coil scenarios.

\begin{table}[!ht]
 	\makeatletter\def\@captype{table}\makeatother\caption{Average (with standard deviation) multi-coil \textbf{reconstruction} results, in terms of SSIM, PSNR and NMSE, under different undersampling patterns. The best and second-best results are marked in {\color{red}red} and {\color{blue}blue}, respectively. $P<$ 0.001 was considered as a statistically significant level.}
 	  \resizebox{0.49\textwidth}{!}{
 	  \setlength\tabcolsep{2pt}
				\begin{tabular}{r||cccc}
			        \hline\thickhline
			        \rowcolor{mygray}
			        {\textbf{Multi-coil}}&  \multicolumn{4}{c}{Reconstruction/Random 4$\times$} \\\hline
			        \rowcolor{mygray}
			     {Method}
			        &~~SSIM$\uparrow$~~ &~~PSNR$\uparrow$~~ &~~NMSE$\downarrow$~~&~~$P$ values\\ \hline\hline
                    SPIRiT~\cite{lustig2010spirit} &0.782$\pm$0.07 &32.1$\pm$0.91 &0.037$\pm$0.02&$<0.001$/$<0.001$/$<0.001$ \\
                    L1-SPIRiT~\cite{murphy2010clinically} &0.830$\pm$0.05 &33.1$\pm$0.72 &0.032$\pm$0.03&$<0.001$/$<0.001$/$<0.001$ \\
                    VN-Net~\cite{12} &0.900$\pm$0.05 &36.0$\pm$0.55 &0.030$\pm$0.03&$<0.001$/$<0.001$/$<0.001$ \\
                    MoDL~\cite{aggarwal2018modl} &{\color{blue}0.902$\pm$0.04} &{\color{blue}36.2$\pm$0.77} &{\color{blue}0.028$\pm$0.02}&$<0.001$/$<0.001$/$<0.001$ \\
                    \textbf{MTrans}   &{\color{red}0.932$\pm$0.03} &{\color{red}37.2$\pm$0.69} &{\color{red}0.022$\pm$0.02}&$-$\\ \hline
                    
                    \hline
	        	\end{tabular}
	        	}
	\captionsetup{font=small}

	\label{multi-coil} 
\end{table}

\subsection{Statistical Analysis}
Aside from the quantitative evaluation of the two tasks, we also performed a statistical significance analysis to prove the effectiveness of our method. Here, following~\cite{ye2019deep}, we used paired Student’s t-test to evaluate the significant difference between the two methods. As can be seen from the $P$ values in Tables~{\color{blue}\ref{table1}} and~{\color{blue}\ref{table2}}, for the p-values that are greater than 0.001, we report the specific values, while those less than 0.001 will not be given specific values. Our results and those given by the comparison methods are statistically different in nearly all cases with p-values smaller than 0.001. For only few cases as the reviewers have mentioned, we observed that the p-values are actually a little bit large than 0.001, but still smaller than 0.05. It is worth noting that our method still has a statistically significant improvement over the various multi-modal baselines. This supports our previous discussion that our method can transfer multi-scale features from the target modality to the auxiliary modality, resulting in higher quality target images, even compared to the currently available state-of-the-art methods.

\subsection{Ablation Studies}
In this section, we first investigate the effectiveness of our approach by comparing it with different fusion strategies. Then, we analyze the trade-off effects between the two modalities. 
\begin{figure}[!t]
\centering
  \includegraphics[width=0.46\textwidth]{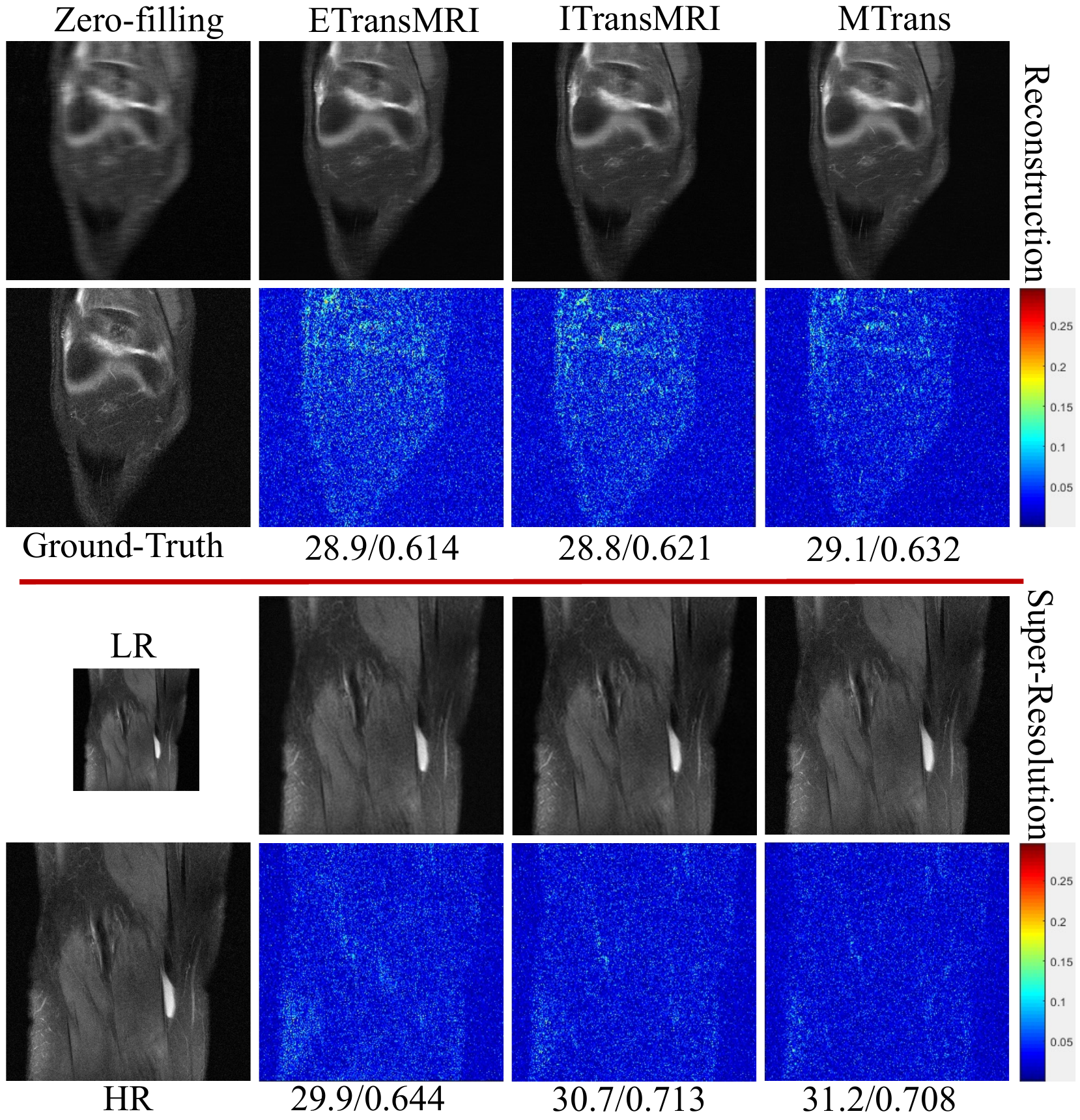}
  \caption{Ablation study of the key components in our method, where ETransMRI and ITransMRI are two variations of our model. The first two and last two rows are the reconstruction and SR results, respectively.}
  \label{figab1}
\end{figure}

\begin{table}[!t]
 \centering
 	\makeatletter\def\@captype{table}\makeatother\caption{Ablation study (with standard deviation) of different fusion strategies in our method regarding the reconstruction results. $P<$ 0.001 was considered as a statistically significant level. The paired Student’s t-test results show statistical difference ($P<$ 0.001).} 
        \label{table3}
			\resizebox{1\linewidth}{!}{
				\setlength\tabcolsep{4pt}
				\renewcommand\arraystretch{1.1}
				\begin{tabular}{r||ccc||ccc}
			        \hline\thickhline
			        \rowcolor{mygray}
			        {{}}&  \multicolumn{3}{c}{\textbf{fastMRI}} &\multicolumn{3}{c} {\textbf{uiMRI}} \\\hline
			        \rowcolor{mygray}

			     {Method}
			        &~~SSIM$\uparrow$~~ &~~PSNR$\uparrow$~~ &~~NMSE$\downarrow$~~&~~SSIM$\uparrow$~~ &~~PSNR$\uparrow$~~ &~~NMSE$\downarrow$~~ \\ \hline\hline
                    ETransMRI  &0.619$\pm$0.03 &29.0$\pm$1.10 &0.037$\pm$0.01 &0.915$\pm$0.02 &30.5$\pm$1.43 &0.030$\pm$0.004\\
                    ITransMRI &0.622$\pm$0.03 &29.2$\pm$0.90 &0.038$\pm$0.02 &0.920$\pm$0.03  &31.0$\pm$1.21 &0.028$\pm$0.005\\ 
                    STransMRI &{\color{blue}0.623$\pm$0.04} &{\color{blue}29.2$\pm$0.87} &{\color{blue}0.037$\pm$0.02} &{\color{blue}0.922$\pm$0.04}  &{\color{blue}31.2$\pm$1.33} &{\color{blue}0.026$\pm$0.004}\\\hline
                    \textbf{MTrans}  &{\color{red}0.628$\pm$0.03} &{\color{red}29.3$\pm$0.89} &{\color{red}0.035$\pm$0.01}&{\color{red}0.931$\pm$0.02} &{\color{red}31.7$\pm$1.33} &{\color{red}0.024$\pm$0.004}\\ \hline

 \end{tabular}}
 \label{ab1}
\end{table}

\begin{table}[!t]
 \centering 
 	\makeatletter\def\@captype{table}\makeatother\caption{Ablation study (with standard deviation) of different fusion strategies in our method regarding the SR results. The paired Student’s t-test results show statistical difference ($P<$ 0.001).} 
        \label{table4}
 			\resizebox{1\linewidth}{!}{
				\setlength\tabcolsep{4pt}
				\renewcommand\arraystretch{1.1}
				\begin{tabular}{r||ccc||ccc}
			        \hline\thickhline
			        \rowcolor{mygray}
			        {{}}&  \multicolumn{3}{c}{\textbf{fastMRI}} &\multicolumn{3}{c} {\textbf{uiMRI}} \\\hline
			        \rowcolor{mygray}

			     {Method}
			        &~~SSIM$\uparrow$~~ &~~PSNR$\uparrow$~~ &~~NMSE$\downarrow$~~&~~SSIM$\uparrow$~~ &~~PSNR$\uparrow$~~ &~~NMSE$\downarrow$~~ \\ \hline\hline
                    ETransMRI &0.668$\pm$0.02 &31.0$\pm$1.21 &0.037$\pm$0.02 &0.953$\pm$0.05 &35.6$\pm$1.09 &0.006$\pm$0.002 \\ 
                    ITransMRI &0.698$\pm$0.03 &31.0$\pm$1.10 &{\color{blue}0.034$\pm$0.03} &0.942$\pm$0.06 &35.8$\pm$1.02 &0.006$\pm$0.004\\ 
                    STransMRI &{\color{blue}0.699$\pm$0.02} &{\color{blue}31.1$\pm$0.90} &0.038$\pm$0.02 &{\color{blue}0.950$\pm$0.05}  &{\color{blue}35.9$\pm$1.21} &{\color{red}0.005$\pm$0.005}\\ \hline
                    \textbf{MTrans}   &{\color{red}0.719$\pm$0.02} &{\color{red}31.9$\pm$1.19} &{\color{red}0.031$\pm$0.02} &{\color{red}0.959$\pm$0.05} &{\color{red}36.1$\pm$0.99} &{\color{red}0.005$\pm$0.003}\\ \hline

 \end{tabular}
}
 \label{ab2}
\end{table}

\begin{table}[!t]
 \centering
 	\makeatletter\def\@captype{table}\makeatother\caption{Ablation study (with standard deviation) on the cross attention module and transformer.The paired Student’s t-test results show statistical difference ($P<$ 0.001).}
        \label{table3}
			\resizebox{1\linewidth}{!}{
				\setlength\tabcolsep{4pt}
				\renewcommand\arraystretch{1.1}
				\begin{tabular}{r||ccc||ccc}
			        \hline\thickhline
			        \rowcolor{mygray}
			        {{}}&  \multicolumn{3}{c}{\textbf{Reconstruction}} &\multicolumn{3}{c} {\textbf{SR}} \\\hline
			        \rowcolor{mygray}

			     {Method}
			        &~~SSIM$\uparrow$~~ &~~PSNR$\uparrow$~~ &~~NMSE$\downarrow$~~&~~SSIM$\uparrow$~~ &~~PSNR$\uparrow$~~ &~~NMSE$\downarrow$~~ \\ \hline\hline
                    $w$/$o$-Trans &0.589$\pm$0.04 &28.0$\pm$1.23 &0.043$\pm$0.02 &0.659$\pm$0.03 &28.6$\pm$1.29 &0.066$\pm$0.07\\ 
                    $w$/$o$-CA &{\color{blue}0.611$\pm$0.02} &{\color{blue}28.7$\pm$0.88} &{\color{blue}0.039$\pm$0.05} &{\color{blue}0.686$\pm$0.05}  &{\color{blue}29.8$\pm$1.33} &{\color{blue}0.049$\pm$0.06}\\ \hline
                    \textbf{MTrans}  &{\color{red}0.628$\pm$0.03} &{\color{red}29.3$\pm$0.89} &{\color{red}0.035$\pm$0.01}&{\color{red}0.719$\pm$0.02} &{\color{red}31.9$\pm$1.19} &{\color{red}0.031$\pm$0.02}\\ \hline
 \end{tabular}
}
 \label{ab5}
\end{table}

\begin{table}[!t]
 \centering
 	\makeatletter\def\@captype{table}\makeatother\caption{Ablation study (with standard deviation) on the CNN-based attention schemes.}

			\resizebox{1\linewidth}{!}{
				\setlength\tabcolsep{4pt}
				\renewcommand\arraystretch{1.1}
				\begin{tabular}{r||ccc||ccc}
			        \hline\thickhline
			        \rowcolor{mygray}
			        {{}}&  \multicolumn{3}{c}{\textbf{Reconstruction}} &\multicolumn{3}{c} {\textbf{SR}} \\\hline
			        \rowcolor{mygray}
			        
			     {Method}
			        &~~SSIM$\uparrow$~~ &~~PSNR$\uparrow$~~ &~~NMSE$\downarrow$~~&~~SSIM$\uparrow$~~ &~~PSNR$\uparrow$~~ &~~NMSE$\downarrow$~~ \\ \hline\hline
                    CNN-SA &{\color{blue}0.612$\pm$0.06} &{\color{blue}28.7$\pm$1.07} &{\color{blue}0.039$\pm$0.05} &0.693$\pm$0.02 &30.7$\pm$0.96 &{\color{blue}0.034$\pm$0.05}\\ 
                    CNN-CSA &0.605$\pm$0.03 &28.6$\pm$1.02 &0.040$\pm$0.07 &{\color{blue}0.698$\pm$0.04}  &{\color{blue}30.8$\pm$1.21} &{\color{blue}0.034$\pm$0.04}\\ \hline
                    \textbf{MTrans}  &{\color{red}0.628$\pm$0.03} &{\color{red}29.3$\pm$0.89} &{\color{red}0.035$\pm$0.01}
                    &{\color{red}0.719$\pm$0.02} &{\color{red}31.9$\pm$1.19} &{\color{red}0.031$\pm$0.02}\\ \hline
 \end{tabular}
}
 \label{ab6}
\end{table}

\subsubsection{Comparison with Joint Reconstruction}
\begin{figure}[t]
\centering
  \includegraphics[width=0.46\textwidth]{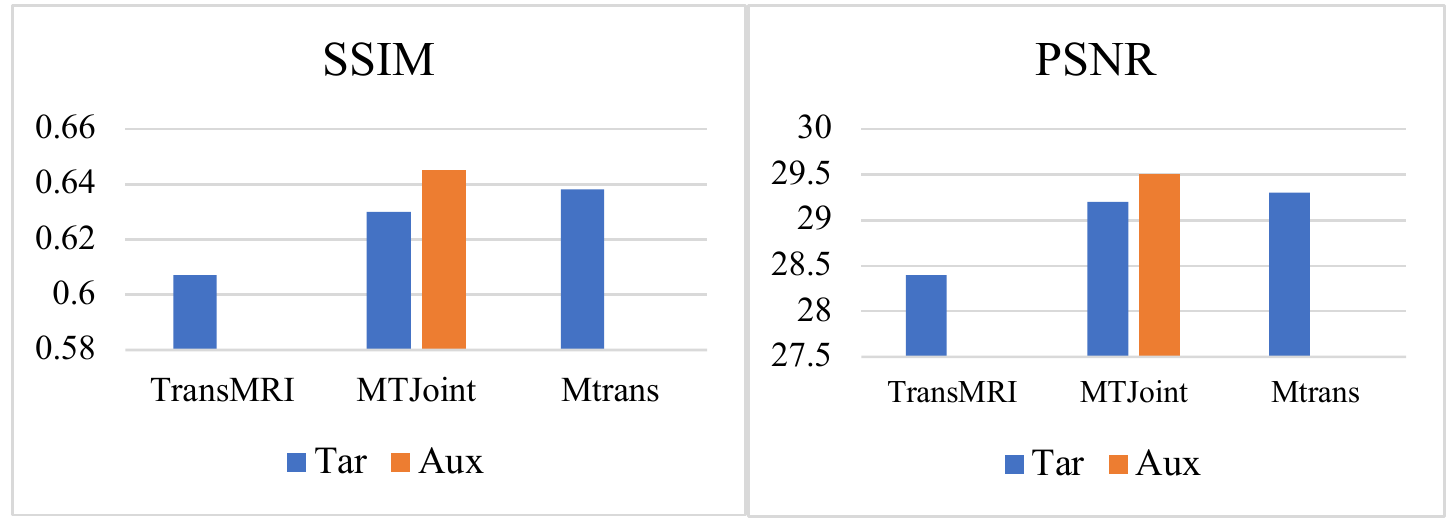}
  \caption{Comparison with the joint reconstruction. MTJoint represents our model, but the input target and auxiliary modalities are both undersampled images, where Tar represents the target modality and Aux represents the auxiliary modality.}
  \label{joint} 
\end{figure}

Here, we test whether our method is still effective when both the auxiliary modality and the target modality need to be reconstructed. We record the reconstruction results under the random undersampling pattern with 4$\times$ acceleration in~\figref{joint}, where MTJoint represents our model, but the input target and auxiliary modalities are both undersampled images. As can be seen from this figure, without the multi-modal fusion mechanism, the PSNR and SSIM results of single-modal TransMRI were the lowest. When both two modalities are undersampled data, MTJoint obtain PSNR = 29.2 dB and SSIM = 0.630 on the target modality, PSNR = 29.3 dB and SSIM = 0.635 on the auxiliary modality, respectively. Note that this result is much higher than the baselines in both single- and multi-modal, see Table~{\color{blue}\ref{table1}}. For the target modality, MTrans provides higher results than MTJoin. This is mainly because the auxiliary modality in MTrans are fully sampled and can provide detailed supplementary information.
However, the results of MTJoint demonstrate that MTrans is a powerful method that can deal with the case that both the auxiliary modality and the target modality are undersampled.

\subsubsection{Comparison of Different Fusion Schemes}
To evaluate the effectiveness of the key components in our model, we study two models in the ablation study. The first is a multi-modal transformer with early fusion, named ETransMRI, where the different modalities are fused as input. ITransMRI is modified from our cross multi-modal transformer, TransMRI. Different from TransMRI, it performs single-scale (large-scale patches in the two branches) fusion with both the auxiliary and target modalities. STransMRI is derived from our method but the two branches use the small-sized image patch.
We summarize the reconstruction and SR results on fastMRI and uiMRI in Table~{\color{blue}\ref{ab1}} and Table~{\color{blue}\ref{ab2}}, respectively. As can be seen, ETransMRI obtains the worst performance, which supports our conclusion that feature-level fusion can provide richer supplementary information than simple fusion. Since early fusion does not learn information at the feature level, it is not the most effective strategy for accelerating multi-modal MR imaging. In addition, because the multi-modal features fused by ITransMRI and STransMRI are with the same size, and different-scale information cannot be extracted between the two modalities, the restoration results of ITransMRI are not the best. The results of STransMRI are lower than MTransMRI. In addition, the number of parameters of STransMRI is larger than that of TransMRI. In contrast, our MTrans inherits the fusion information of different modalities at multiple scales, enhances the fusion features, and captures both high-level context features and local details. The results in this section also demonstrate its strong ability to mine key information for guiding the target modality.

To qualitatively analyze the different fusion schemes, we show visual results on fastMRI with error maps in~\figref{figab1}. The first two are the reconstruction results, while the last two rows are the SR results. From this figure, we can see that both ITransMRI and ETransMRI can effectively restore the image. However, MTrans achieves the lowest texture error, and provides results that are almost as clear as the ground truth. This indicates that our cross attention in the multi-modal transformer is effective for accelerating multi-modal MR imaging in both reconstruction and super-resolution.

\begin{figure}[!t]
\centering
  \includegraphics[width=0.47\textwidth]{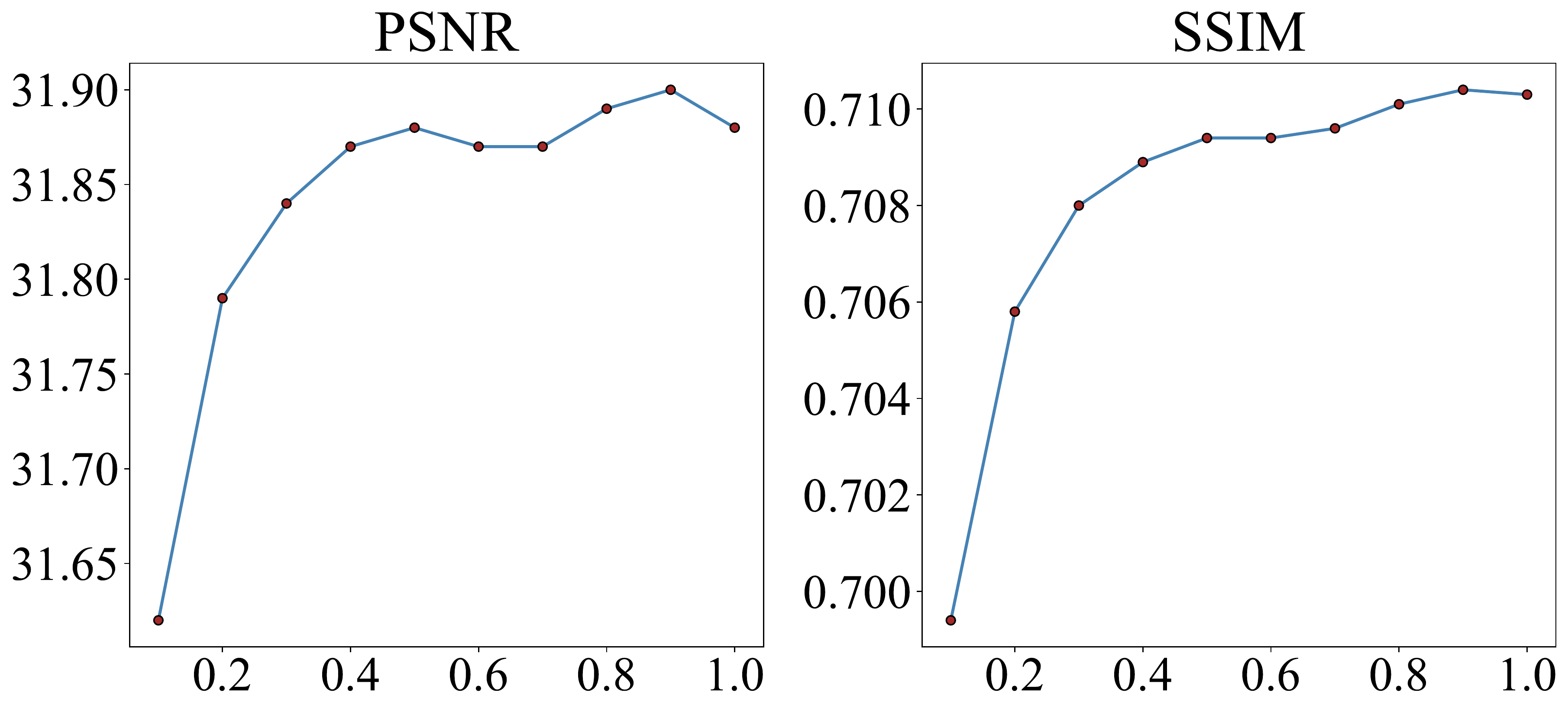}
  \put(-57,-4){\footnotesize $\alpha$}
  \put(-182,-4){\footnotesize $\alpha$}
  \caption{Analysis of trade-off ($\alpha$) between the two modalities in terms of PSNR and SSIM. The greater the value of $\alpha$, the greater the influence of the target modality, and the smaller the influence of the auxiliary modality.}
  \label{alpha}
\end{figure}

\subsubsection{Comparison with CNN-based Attention Schemes}
To investigate the limitations of CNN, we compared our MTrans with self-attention (CNN-SA) and channel-spatial-attention (CNN-CSA) mechanisms, which are based on CNN~\cite{hu2018squeeze}. As can be seen from Table~{\color{blue}\ref{ab6}}, with the help of the self-attention and cross attention, the results of CNN-SA and CNN-CSA is still lower than that of MTrans in both reconstruction and SR. Because the transformer also uses the multi-head attention rather than calculating just one self-attention, this allows the model to deal with different presentation sub-spaces at different locations. Therefore, compared with CNN-SA and CNN-CSA, transformer-based method has stronger capabilities in modeling long-range dependency.

\subsubsection{Baseline Comparisons with Same Loss}

\begin{table}[!ht]
 	\makeatletter\def\@captype{table}\makeatother\caption{Ablation study (with standard deviation) on the various baselines with same loss. The best and second-best results are marked in {\color{red}red} and {\color{blue}blue}, respectively. The paired Student’s t-test results show statistical difference ($P<$ 0.001).}
 	  \resizebox{0.49\textwidth}{!}{
 	  \setlength\tabcolsep{5pt}
				\begin{tabular}{r||ccc}
			        \hline\thickhline
			        \rowcolor{mygray}
			        {\textbf{fastMRI}}&\multicolumn{3}{c}{Reconstruction/Random 4$\times$} \\\hline
			        \rowcolor{mygray}
			     {Method}
			        &~~SSIM$\uparrow$~~ &~~PSNR$\uparrow$~~ &~~NMSE$\downarrow$~~\\ \hline\hline
                    LORAKS~\cite{haldar2013low}  &0.532$\pm$0.08 &25.8$\pm$1.17 &0.060$\pm$0.03 \\
                    MoDL~\cite{aggarwal2018modl} &0.580$\pm$0.05 &27.6$\pm$0.89 &0.049$\pm$0.02 \\
                    UNet~\cite{zbontar2018fastmri} &0.565$\pm$0.06 &28.2$\pm$1.15 &0.046$\pm$0.02\\ 
                    TransMRI~\cite{vaswani2017attention}  &0.607$\pm$0.05 &28.4$\pm$0.89 &0.038$\pm$0.01 \\ 
                    Transmed~\cite{dai2021transmed} &0.609$\pm$0.06 &28.4$\pm$0.99 &0.040$\pm$0.03 \\
                    HyperDense-Net~\cite{dolz2018hyperdense} &0.600$\pm$0.04 &28.3$\pm$1.00 &0.038$\pm$0.02 \\
                    MDUNet~\cite{xiang2018deep} &0.592$\pm$0.06 &28.4$\pm$1.12 &0.053$\pm$0.02\\
                    rsGAN~\cite{dar2020prior} &{\color{blue}0.620$\pm$0.07} &{\color{blue}29.0$\pm$1.12} &{\color{blue}0.035$\pm$0.03}\\ 
                    \textbf{MTrans}   &{\color{red}0.638$\pm$0.03} &{\color{red}29.3$\pm$0.89} &{\color{red}0.030$\pm$0.01}\\ \hline
                    \rowcolor{mygray}
                    {\textbf{fastMRI}}&  \multicolumn{3}{c}{SR 4$\times$} \\\hline
                    EDSR~\cite{lim2017enhanced} &0.580$\pm$0.04 &28.1$\pm$1.64 &0.045$\pm$0.04\\ 
                    TransMRI~\cite{vaswani2017attention} &0.600$\pm$0.03 &29.9$\pm$1.44 &0.048$\pm$0.02\\ 
                    Transmed~\cite{dai2021transmed} &0.673$\pm$0.05 &29.4$\pm$1.35 &0.053$\pm$0.05\\
                    HyperDense-Net~\cite{dolz2018hyperdense} &0.640$\pm$0.04 &30.4$\pm$1.62 &0.042$\pm$0.04\\
                    PRO~\cite{lyu2020multi} &{\color{blue}0.710$\pm$0.03} &{\color{blue}31.0$\pm$1.32} &{\color{blue}0.032$\pm$0.04}\\ 
                    MCSR~\cite{zeng2018simultaneous} &0.690$\pm$0.02 &30.5$\pm$1.22 &0.040$\pm$0.04\\
                    \textbf{MTrans}   &{\color{red}0.719$\pm$0.02} &{\color{red}31.9$\pm$1.19} &{\color{red}0.031$\pm$0.02}\\ \hline
                    
                    \hline
	        	\end{tabular}
	        	}
	\captionsetup{font=small}

	\label{ab7} 
\end{table}

 To demonstrate the benefits owe to our network architecture rather than the loss functions, we modify all the baselines to use the same loss function as our method. In Table~{\color{blue}\ref{ab7}}, we report the reconstruction results under the random sampling pattern with 4$\times$ acceleration and SR results with 4$\times$ enlargement. For the multi-modal classification methods HyperDense-Net~\cite{dolz2018hyperdense} and Transmed~\cite{dai2021transmed}, we have already changed the classification loss to $L_1$ loss in Table~{\color{blue}\ref{table1}}. For UNet~\cite{zbontar2018fastmri}, TransMRI~\cite{vaswani2017attention}, and EDSR~\cite{lim2017enhanced}, the original supervised loss is $L_1$ loss. Therefore, we retained the other methods with the same loss as ours. For rsGAN~\cite{dar2020prior} and PRO~\cite{lyu2020multi}, we remove the discriminators including the perceptual loss and adversarial loss, and use $L_1$ as the supervised loss. For LORAKS~\cite{haldar2013low}, MoDL~\cite{aggarwal2018modl}, MDUNet~\cite{xiang2018deep}, and MCSR~\cite{zeng2018simultaneous}, we change the MSE loss as $L_1$ loss. As can be seen from this table, without the discriminator, the PSNR and SSIM results of both rsGAN~\cite{dar2020prior} and PRO~\cite{lyu2020multi} are improved. This is consistent with the previous studies that GAN-based structures tend to produce low PSNR but good visual effects~\cite{blau2018perception}. When we change the MSE loss of LORAKS~\cite{haldar2013low}, MoDL~\cite{aggarwal2018modl}, MDUNet~\cite{xiang2018deep}, and MCSR~\cite{zeng2018simultaneous} to $L_1$ loss, their results decreased slightly. This might be because MSE loss maximizes PSNR and has better convergence than $L_1$ loss~\cite{lim2017enhanced}.

\subsubsection{Strength of the CA Module in Transformer}
To understand the strength of the cross attention (CA) module in Transformer, we investigate ablation study on the heads and tails without transformer ($w$/$o$-Trans), transformer without the CA module ($w$/$o$-CA). We summarize the reconstruction and SR results on fastMRI in Table~{\color{blue}\ref{ab5}}. As can be seen, both the reconstruction and SR performance drops dramatically without transformer, because it cannot deal with long-range relationships well. The results of $w$/$o$-CA and MTrans show that CA module plays an important role in the fusion between the different modalities.

\subsubsection{Parameter Analysis}

Here, we analyze the number of parameters for each method. Specifically, the parameters of heads, tails, and the multi-modal transformer module in our method are 4.78KB, 0.02KB, and 189.2M, respectively. UNet requires 31M parameters. HyperDense-Net and MDUNet only needed 11M and 10M parameters, respectively. EDSR requires 43M parameters. MCSR is composed of two EDSR frameworks and requires 86M parameters. PRO adopts a progressive model which requires 7M parameters and rsGAN requires 62M parameters. Transmed is built on the transformer framework, therefore the number of parameters is larger than other methods, \ie, it requires 145M parameters. Although our MTrans built on transformer requires the largest number of parameters, the results in Table~{\color{blue}\ref{table1}}, Table~{\color{blue}\ref{table2}}, and Table~{\color{blue}\ref{multi-coil}} show that our MTrans provides the best performance in both MRI reconstruction and SR. In the future, we will seek more efficient solutions to reduce the memory cost of our model. In our experiments, the number of cross-transformer encoders is set to 4. The results will drop slightly if the number of transformer encoders or cross-attention heads is reduced. But the results are still higher than all baseline methods. The current number of encoders is the trade-off between network parameters and reconstruction accuracy.

\subsubsection{Effect of Trade-Off Between the Two Modalities}
We next investigate the influence of $\alpha$, which weights the trade-off between the two modalities. Specifically, the ratio $\alpha$ determines the weights of both the target and auxiliary branches. The greater the value of $\alpha$, the greater the influence of the target modality, and the smaller the influence of the auxiliary modality. We record the SR results of our method on fastMRI in~\figref{alpha}. As can be seen, our model achieves the best PSNR and SSIM scores at $\alpha$ = 0.9. When $\alpha$ = 1, the PSNR performance quickly degrades, while the SSIM degrades only slightly. This is likely because the auxiliary modality is fused with the ground truth at multiple scales, so the weight of the auxiliary branch loss is not affected very much.

\subsubsection{Effect of the Number of $N$}

\begin{figure}[!t]
\centering
  \includegraphics[width=0.49\textwidth]{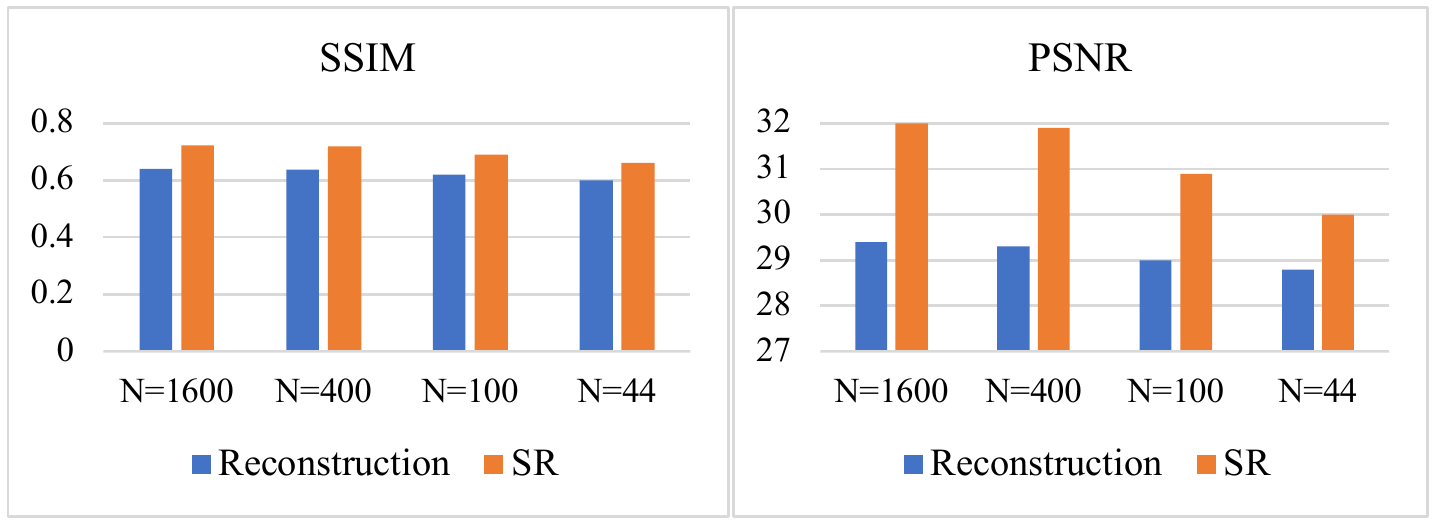}
  \caption{Analysis of the effect of the $N$ values on the fastMRI.}
  \label{figN}
\end{figure}

The values of $N$ represent the number of patches that the input image has been cropped. To verify whether the values of $N$ are important to our method, we record the reconstruction results under random undersampling pattern with 4$\times$ acceleration and SR results with 4$\times$ enlargement results of various $N$ values on fastMRI in~\figref{figN}. As can be seen, the smaller the value of $N$, the smaller PSNR and SSIM results.
However, when $N$ is greater than $400$, the PSNR and SSIM improvement is marginal.
It should be noted that the parameters of $N=1600$ will be much larger than the parameters of $N=400$, so we set $N=400$ in our experiment.

\section{Conclusion}\label{sec:conclusion}
In this work, we focus on exploring rich global knowledge in image for accelerated multi-modal MR imaging. For this purpose, we proposed a unified transformer framework, named MTrans, for accelerated multi-modal MR imaging, which can be used for MR image reconstruction and SR to effectively restore the target modality under the guidance of the auxiliary modality. By fusing the feature maps of different modalities, the proposed MTrans is helpful to learn the global information of multi-modal MR image, obtaining higher quality reconstructed images and significantly reduce artifacts. In particular, the proposed cross attention module can explore the fusion strategy in different scales, which provides both obtain high-level context features and local details. We conducted extensive experiments on the fastMRI and real-world clinical datasets under different settings of undersampling patterns. The results demonstrated our model against outperforms state-of-the-art methods in accelerated MR imaging. This work provides promising guidelines for further research into multi-modal MR imaging with transformers. Although our MTrans provides excellent results on accelerated MR imaging, it still has some shortcomings. For example, MTrans tends to require a large number of parameters than CNN-based methods. Therefore, we will consider to reduce the number of parameters for our transformer-based framework in the future.

{\small
\bibliographystyle{IEEEtran}
\bibliography{ref}
}

\end{document}